\documentclass[]{elsart}
\usepackage{amssymb}
\usepackage{pst-node,pstricks,graphicx}

\begin{document}                % INITIALIZE - DONT CHANGE
% \flushright{Preliminary Draft!}
\begin{frontmatter}
\title{
Characteristics of geomagnetic cascading of ultra-high energy photons
at the southern and northern sites of the Pierre Auger Observatory
}

\author[a]{P.~Homola,\corauthref{cor1}}
\author[a,b]{M.~Risse,}
\author[b]{R.~Engel,}
%\author[]{...}
 \author[a,b]{D.~G\'ora,}
 \author[a]{J.~P\c{e}kala,}
 \author[a]{B.~Wilczy\'nska,}
 \author{and}
 \author[a]{H.~Wilczy\'nski}
\corauth[cor1]{ {\it Correspondence to}: P.~Homola
(piotr.homola@ifj.edu.pl)}
\address[a]{
H.~Niewodnicza\'nski Institute of Nuclear Physics, Polish Academy of Science,
ul.~Radzikowskiego 152,
31-342 Krak\'ow, Poland
}
\address[b]{
Forschungszentrum Karlsruhe, Institut f\"ur Kernphysik, 76021 Karlsruhe, Germany
}
% \linenumbers
%%%%%%%%%%%%%%%%%%%%%%%%%%%%%%%%%%%%%%%%%%%%%%%%%%%%%%%%%%%%%%%%%%%%%
\begin{abstract}                % DON'T CHANGE THIS LINE

Cosmic-ray photons above 10$^{19}$~eV can convert in the geomagnetic
field and initiate a pre\-shower, i.e.~a particle cascade
before entering the atmosphere.
% The characteristics of this process depend strongly on
% the photon trajectory through the magnetic field.
% This leads to significant differences in the preshower
% characteristics for experimental sites located at
% different geographical positions.
We compare the preshower characteristics at the
southern and northern sites of
the Pierre Auger Observatory.
In addition to a shift of the preshower patterns on the
sky due to the different pointing of the local magnetic
field vectors, the fact that the northern Auger site
is closer to the geomagnetic pole results in a different energy
dependence of the preshower effect: photon conversion can start
at smaller energies, but 
large conversion probabilities ($>90\%$) are reached
for the whole sky at higher energies compared to the
southern Auger site.
We show how the complementary preshower features
at the two sites can be used to search for 
ultra-high energy photons among cosmic rays.
In particular, the different preshower characteristics
at the northern Auger site may provide an elegant and
unambiguous confirmation if a photon signal is
detected at the southern site.

\end{abstract}
\end{frontmatter}
% \linenumbers
%%%%%%%%%%%%%%%%%%%%%%%%%%%%%%%%%%%%%%%%%%%%%%%%%%%%%%%%%%%%%%%%%%%%%
% 11111111111111111111
\section{Introduction}
\label{sec-intro}
Substantial fluxes of cosmic-ray photons at ultra-high energy (UHE),
above 10$^{19}$~eV, are predicted by non-acceleration (top-down) models
of cosmic-ray origin (for example, see~\cite{bhat-sigl}).
At a smaller level, UHE photons are also expected to be produced
in acceleration (bottom-up) models~\cite{models}.
UHE photons can be detected through the particle cascades they initiate
in the atmosphere.
So far, upper limits on the photon flux were 
set~\cite{ave,shinozaki,risse05,troitsky,augerphoton}.
The large exposure expected to be collected during the next years,
in particular by the Pierre Auger Observatory~\cite{auger},
will enormously increase the sensitivity for detecting
UHE photons~\cite{augerphoton}.

Contrary to the case of hadron primaries, UHE photons around
10$^{20}$~eV can interact with the magnetic field of the Earth
before entering the atmosphere~\cite{mcbreen}.
This process is commonly called geomagnetic cascading or preshower
and leads to
a dramatic change of the air shower development for primary
photons~\cite{mcbreen,collection,stanev97,bednarek99,bertou00,homola}.

The probability of magnetic $e^+e^-$ pair production (``photon conversion'') and, in case
of conversion, the synchrotron emission by the produced electrons
depend on the particle energy and on the transverse
component of the local magnetic field~\cite{erber,homola}.
Photon conversion can occur at altitudes up to several thousand kilometers
above the ground, i.e.~well above the atmosphere ($\sim$100~km altitude).
A preshower consists mostly of the secondary photons emitted by the
electrons. A secondary photon of sufficiently high energy may additionally convert in the
magnetic field and produce another electron pair.
The number of preshower particles depends on the specific conditions
and can fluctuate strongly; it can range from a few particles to a few thousand.
Nevertheless, as we demonstrate in this paper, the preshower formation
can be fairly well inferred from characteristics of the observed extensive air shower.
The spread of the preshower particles in transverse distance from the shower
axis and in arrival time at the atmosphere is well below the resolution of
current shower experiments~\cite{homola}.
Hence, also in case of geomagnetic cascading, the subsequent air shower
would be observed as one event by the experiments.

Since the probability of photon conversion and the synchrotron emission
spectrum depend on the local transverse magnetic field,
characteristics of the preshower process
(such as conversion probability, or the energy spectrum of the
preshower particles)
depend strongly on the specific trajectory through the magnetosphere.
For a given experimental site, this leads to a large directional
dependence of preshower characteristics within the local coordinate system of
zenith $\theta$ and azimuth $\phi$.
In addition, as already noted in the work of McBreen and Lambert
25 years ago~\cite{mcbreen}, a dependence is expected on the location
of the experiment:
experimental sites at different geomagnetic latitudes have
different local magnetic field conditions, which can affect 
preshower characteristics for UHE photons.

The comparison of preshower characteristics of UHE photons expected at
two sites located at different hemispheres of the Earth is of particular
current interest due to the construction of the
Pierre Auger Observatory.
The southern part of the Observatory (``Auger South'') is situated near
Malargue (Argentina) at 69.2$^\circ$~W, 35.2$^\circ$~S.
The construction of Auger South is approaching
completion and first results were obtained, including an upper limit
to UHE photons~\cite{augerphoton}.
To achieve full sky coverage, the northern site (``Auger North'') is
planned in Colorado (USA) at 102.7$^\circ$~W, 37.7$^\circ$~N.
At Auger South, the magnetic field of $\sim$24.6~$\mu T$
points upward to $\theta \sim 55^\circ ,\phi \sim 87^\circ$.
At Auger North, the magnetic field of $\sim$52.5~$\mu T$
points downward from 
$\theta \sim 25^\circ ,\phi \sim 262^\circ$.\footnote{Azimuth
is defined in this work counterclockwise from geographic East.
For instance, $\phi = 0^\circ$ means East, $\phi = 90^\circ$ North etc.}

Previous works in the literature focused rather on the directional
dependence of preshower characteristics of UHE photons for one
given site than comparing the conditions at different sites.
Examples of magnetic field strengths at various sites were
given in~\cite{stanev97}. In that work, the observable of shower
size (total number of electrons at ground) was studied.
It was noted that azimuth regions of smaller (or stronger)
preshower effect can be identified that may differ between the sites.
In~\cite{bednarek99}, the altitude dependence of the transverse
component of the magnetic field was compared for the two Auger sites
for a few directions, and preshower particle spectra were studied.
E.g.~in~\cite{bertou00}, it was pointed out that at Auger North,
due to the larger magnetic field, the preshower effect starts
at a lower energy. A particular focus of that work was the study
of observables for the ground array, where differences in the
detector signals for converted and unconverted primary photons
were found.

In this work, making use of our PRESHOWER code~\cite{homola}
linked to CONEX, a fast shower simulation program~\cite{conex},
we perform a dedicated
comparison of the preshower characteristics of UHE photons for the
conditions of the Auger North and Auger South sites. The geomagnetic field in 
PRESHOWER is described by the IGRF model~\cite{igrf,tsygan}.  
We find that the transition range in energy from small to large conversion
probabilities is larger at Auger North, starting at smaller
but ending at higher energy.
We study how the differences of the preshower characteristics
between the two sites can be used to perform a complementary search
for UHE photons.
In particular, the situation may arise where a possible UHE photon
signal detected at Auger South requires an independent cross-check.
As will be shown, such a cross-check could come from Auger North.
In fact, making use of the UHE preshower effect by observing from
a different experimental site seems the most elegant and natural way
for cross-checking an UHE photon signal claim.

While the study is performed for the specific case of the two
Auger sites, the general findings hold for any two sites with
sufficiently different local magnetic field conditions.
The results shown for Auger North hold, to a good extend, also
for the HiRes~\cite{hires} and Telescope Array~\cite{telarray}
sites, both of them being located in Utah.

The plan of the paper is as follows.
In Section~\ref{sec-preshower}, preshower features are discussed,
and in Section~\ref{sec-eas}, the corresponding
air shower features are studied.
In Section~\ref{sec-scenarios}, UHE photon scenarios and their
observation at the two different sites are investigated. 
In particular, it is worked out how observations at the two sites
can complement each other.
Conclusions are given in Section~\ref{sec-conclusion}.
In Appendix~\ref{app-a} it is shown how the transverse component of the geomagnetic 
field in the IGRF model varies along different particle trajectories.
In Appendix~\ref{app-b} we show a collection of conversion probability
sky maps supplementing Section \ref{subsec-convprob}.

%%%%%%%%%%%%%%%%%%%%%%%%%%%%%%%%%%%%%%%%%%%%%%%%%%%%%%%%%%%%%%%%%%%%%
% 22222222222222222222

\section{Comparison of preshower features at Auger South and North}
\label{sec-preshower}

\subsection{Conversion probability}
\label{subsec-convprob}
A key parameter to characterize the fate of an UHE photon in the
Earth's magnetic field is the conversion probability $P_{\rm conv}$.
Given the local differential probability of a photon to convert into
an electron pair, $P_{\rm conv}$ results from an integration along the
particle trajectory.
Small values of $P_{\rm conv}$ indicate a large probability of the
UHE photon to enter the atmosphere without conversion and to keep
its original identity. In turn, UHE photons would almost surely
undergo geomagnetic cascading for  values of $P_{\rm conv}$ close to
unity. 

$P_{\rm conv}$ depends on the experimental site, the photon energy,
and the direction of the particle trajectory in the
local coordinate system of zenith $\theta$ and azimuth $\phi$,
$P_{\rm conv}=f($site,$E,\theta ,\phi$).
Thus, for a chosen site and a fixed primary photon energy, sky maps
within the local coordinate system $P_{\rm conv}=f(\theta ,\phi$)
can be produced to study the pattern of UHE photon conversion.

\begin{figure}[t]
\begin{center}
\includegraphics[height=4.8cm,angle=0]{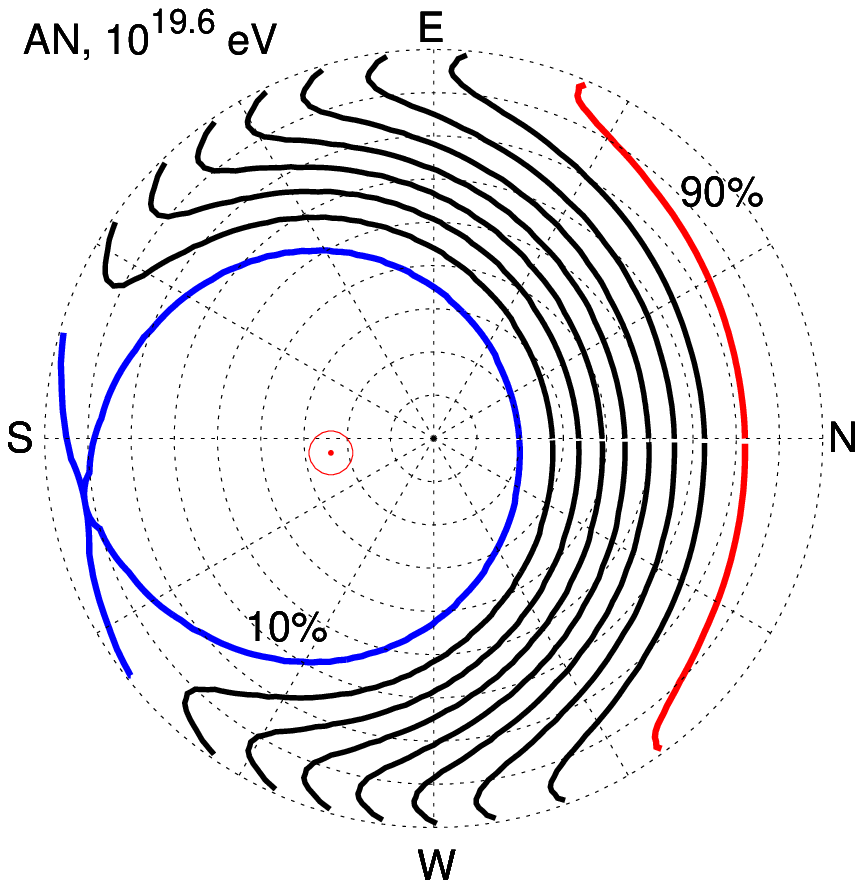}
\includegraphics[height=4.8cm,angle=0]{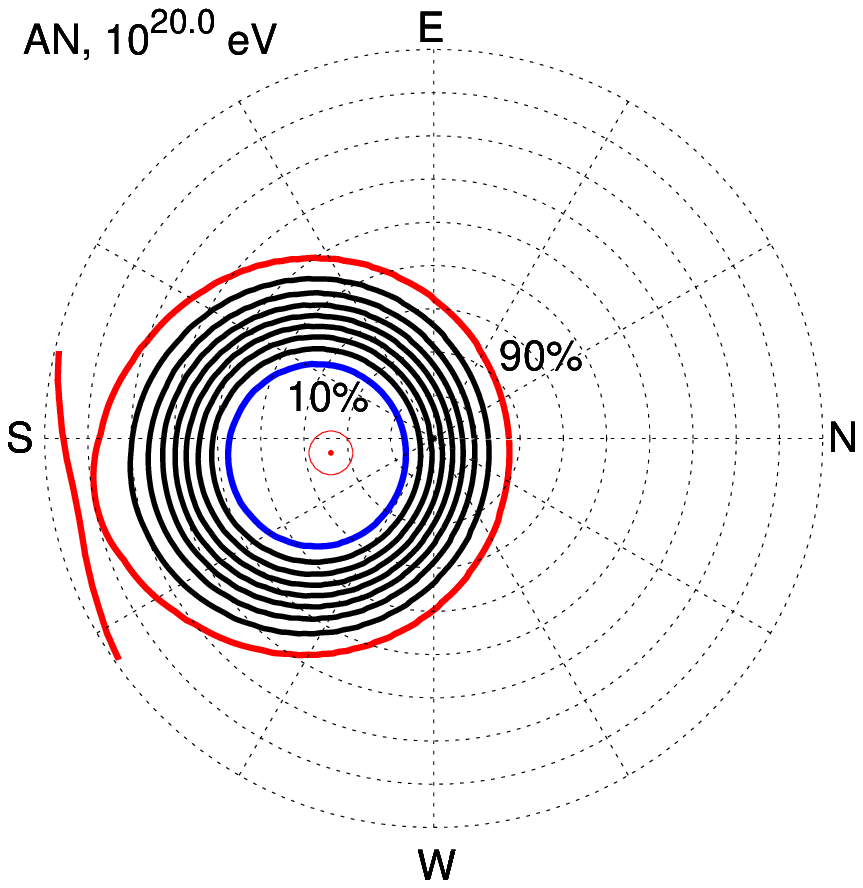}
\includegraphics[height=4.8cm,angle=0]{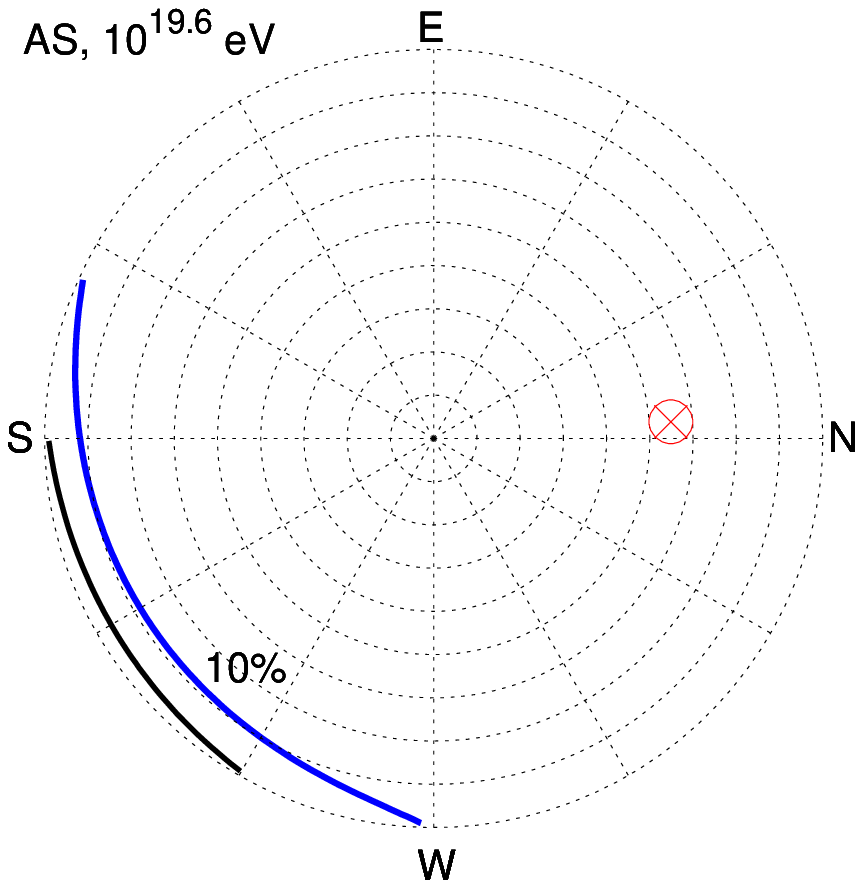}
\includegraphics[height=4.8cm,angle=0]{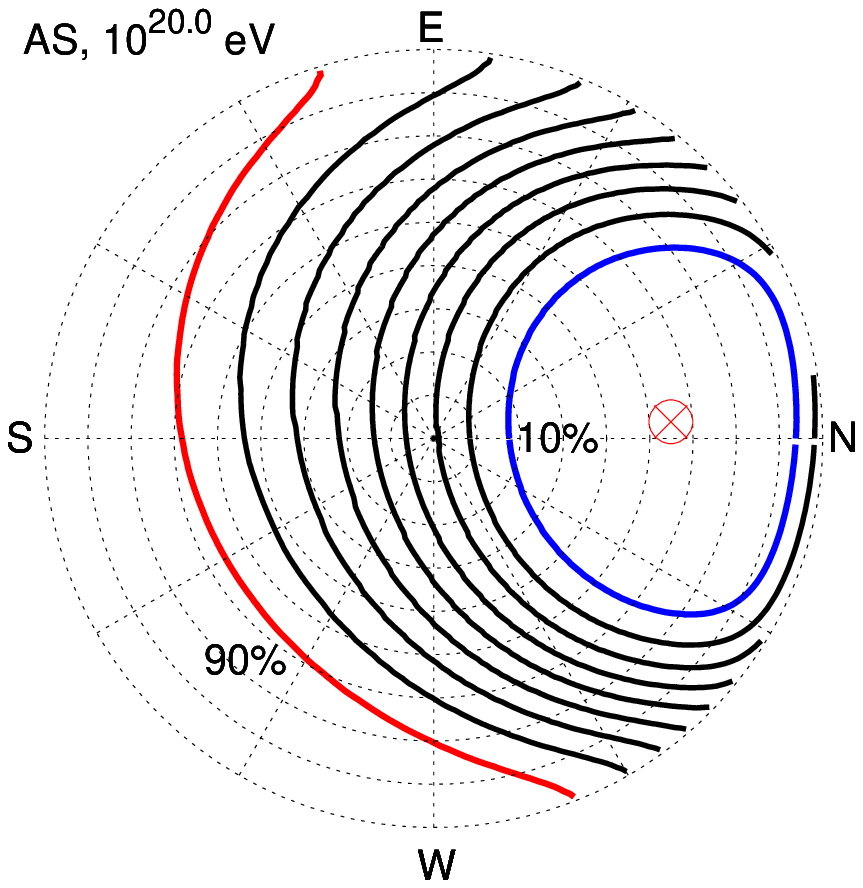}
\includegraphics[height=4.8cm,angle=0]{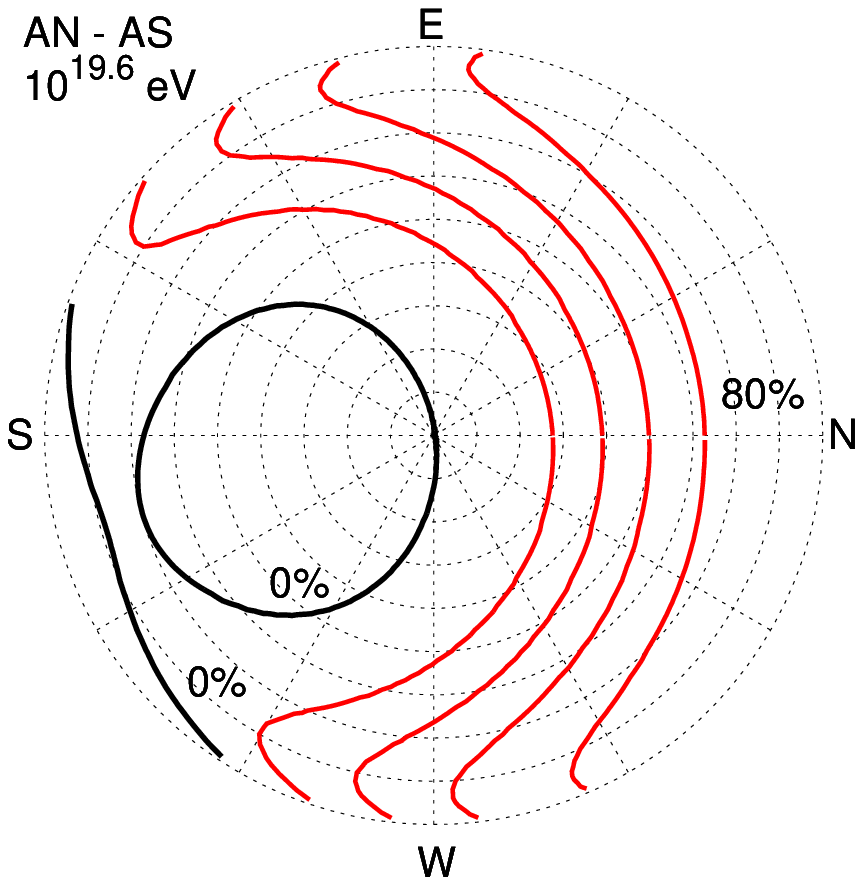}
\includegraphics[height=4.8cm,angle=0]{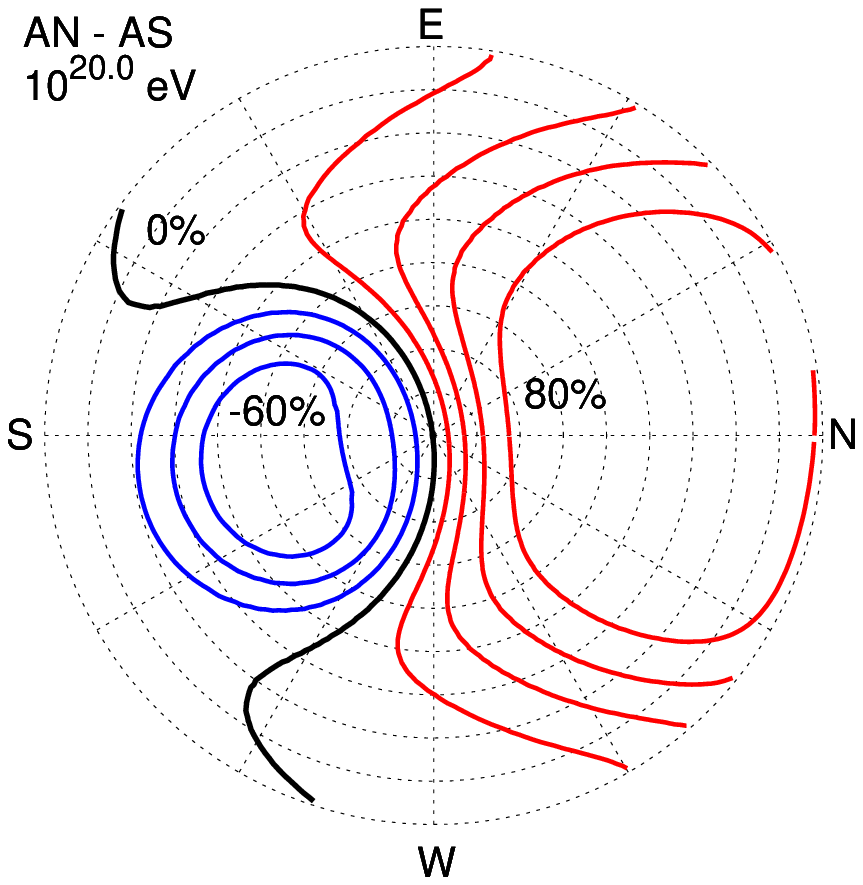}
\caption{
Sky maps of conversion probabilities at $\lg(E/$eV)=19.6 (left)
and 20.0 (right) for Auger North (AN, top), Auger South (AS, middle), and
the difference (AN$-$AS, bottom). Contour lines are given
for the conversion probability
(top and middle rows of plots) with a stepsize of 10\%,
and for the difference of conversion probabilities (bottom plots)
with a stepsize of 20\%.
Azimuthal directions are labeled (``E'' for East etc.).
Zenith angles are given as concentric
circles of 10$^\circ$ steps ($\theta = 0^\circ$ in the center). 
The pointing direction of the local magnetic field vector at
ground is indicated for a specific site (top and middle plots).
}
\label{fig-pc}
\end{center}
\end{figure}

As an example, in Figure~\ref{fig-pc} sky maps are shown
for two different energies $\lg(E/$eV)=19.6 and 20.0,
corresponding to $\sim$ 40~EeV and 100~EeV, both for
Auger North and South (sky maps at various
energies between $\lg(E/$eV)=19.4 and 20.4 are collected
in Appendix~\ref{app-b}).
One can see that at 40~EeV, photon conversion is already
important at Auger North for a large part of the sky.
In contrast to this, at Auger South conversion is very unlikely
for 40~EeV photons independent of their arrival direction.

\begin{figure}[t]
\begin{center}
\includegraphics[height=4.8cm,angle=0]{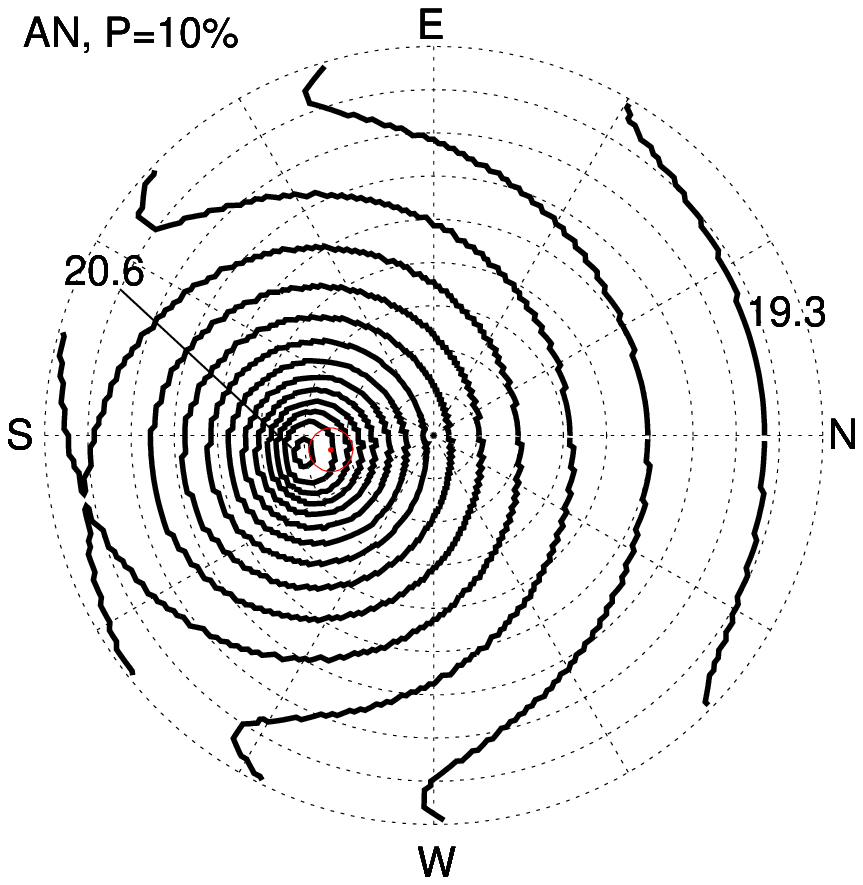}
\includegraphics[height=4.8cm,angle=0]{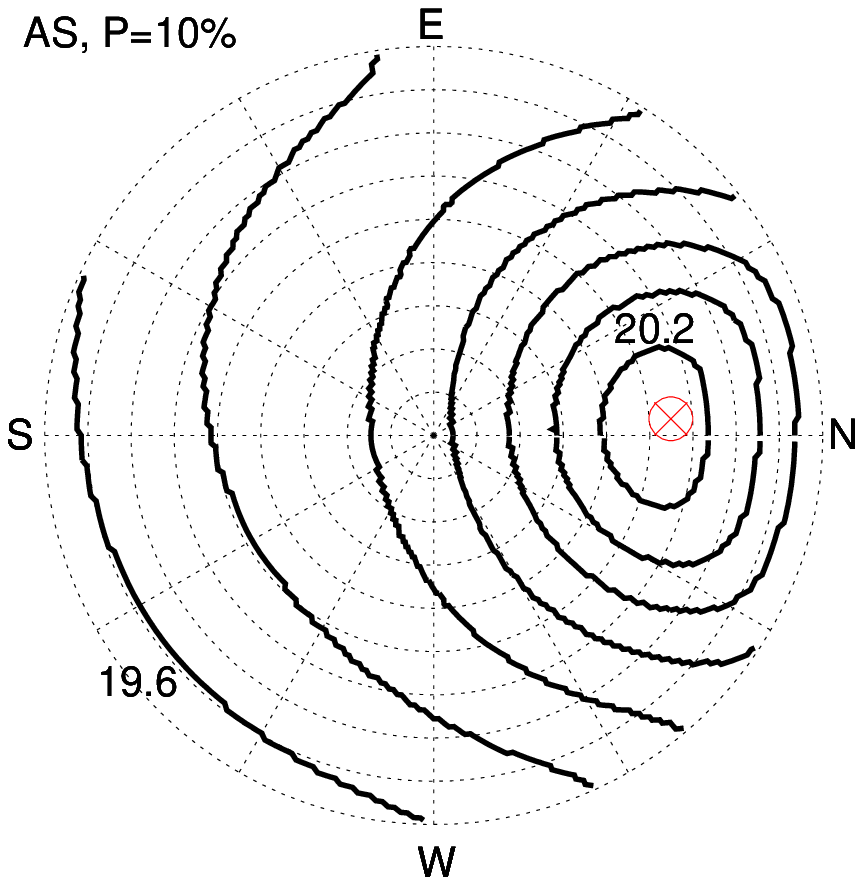}
\includegraphics[height=4.8cm,angle=0]{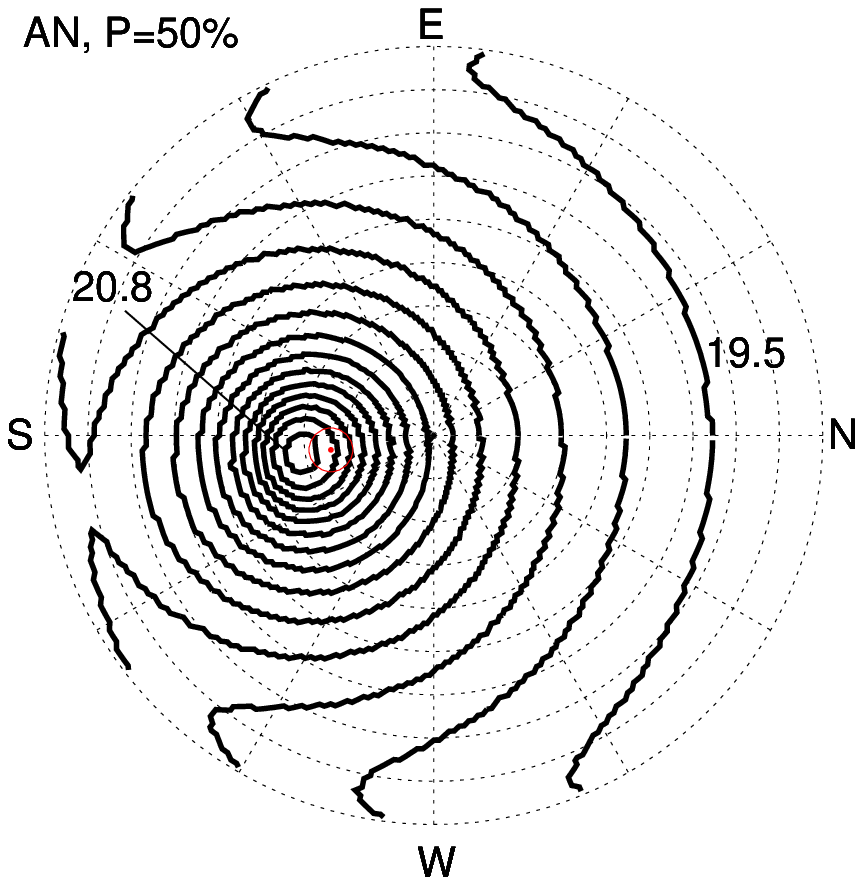}
\includegraphics[height=4.8cm,angle=0]{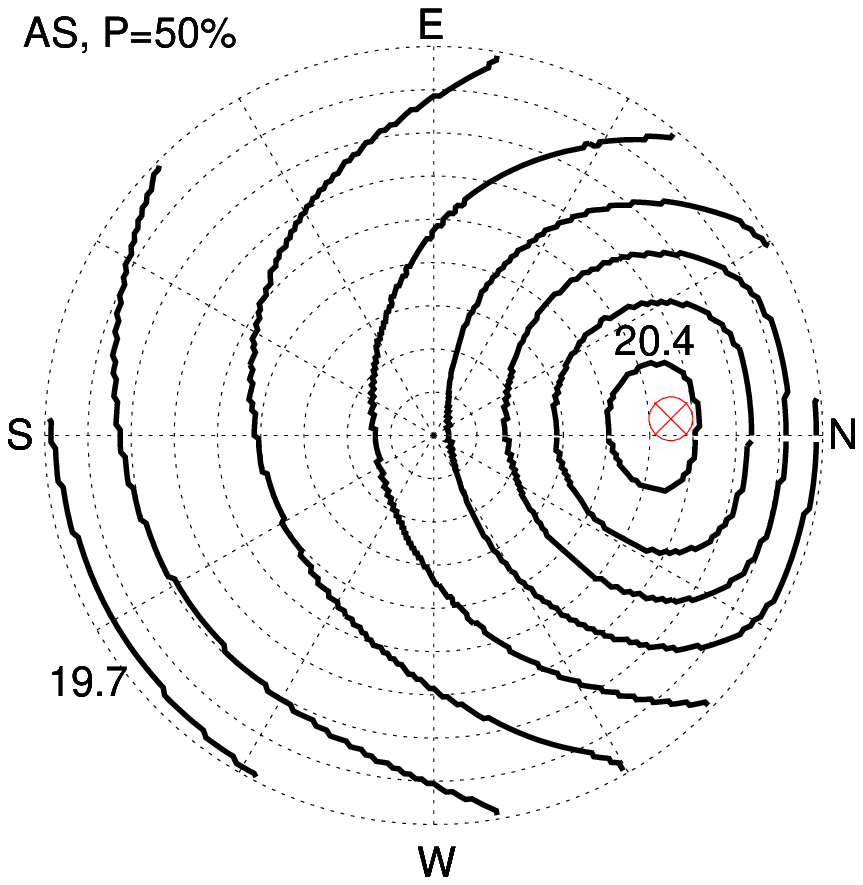}
\includegraphics[height=4.8cm,angle=0]{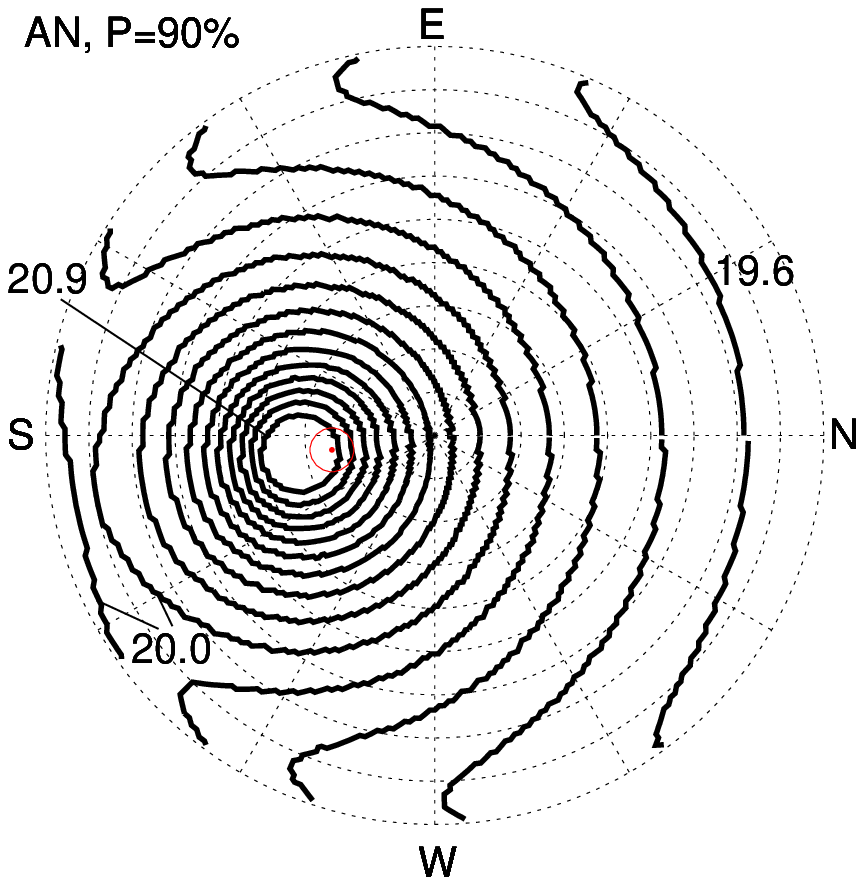}
\includegraphics[height=4.8cm,angle=0]{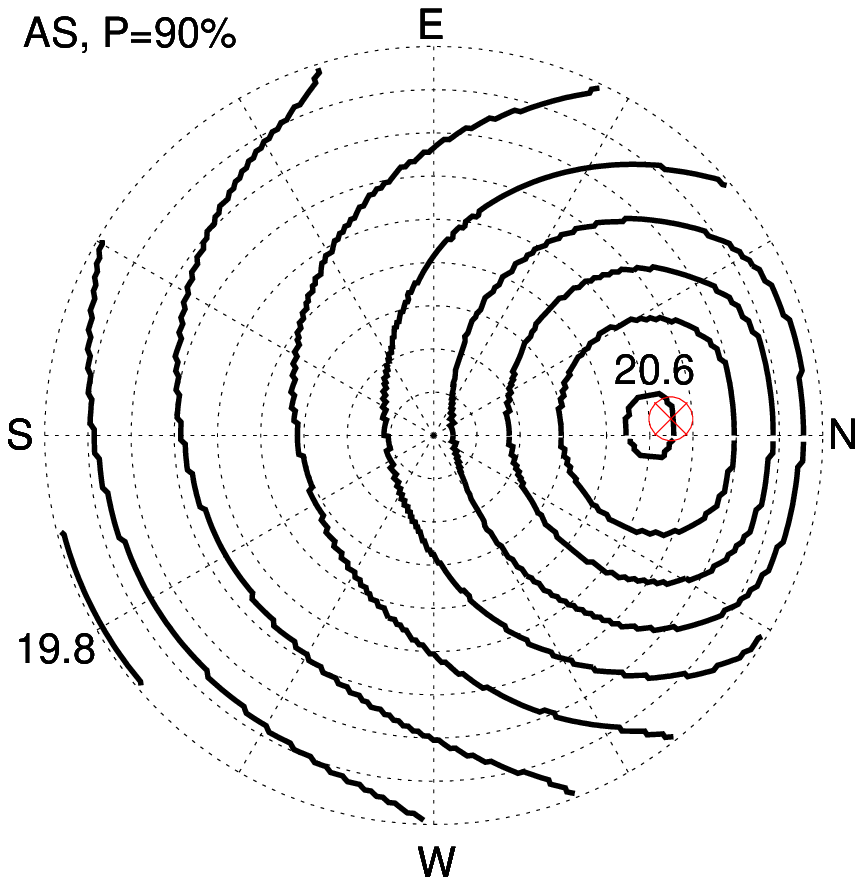}
\caption{
Sky maps of photon energies for fixed conversion probabilities
of 10\% (top), 50\% (middle) and 90\% (bottom)
for Auger North (left) and Auger South (right).
Contour lines are given with a stepsize of 
$\Delta \lg (E/$eV) = 0.1. Values of the minimum and maximum
energies are assigned to the lines (in $\lg (E/$eV)).
See caption of Figure~\ref{fig-pc} for further explanations.
}
\label{fig-pofce}
\end{center}
\end{figure}

The energy above which photon conversion becomes important, 
is expected to be smaller at Auger North due to the larger
local magnetic field strength.
To quantify this, we show in Figure~\ref{fig-pofce} a sky map of photon energies
for a fixed conversion probability of $P_{\rm conv} = 10\%$.
It can be seen that photon conversion starts
at a factor $\sim$2
higher energy at Auger South compared to Auger North.

At energies of 100~EeV, also for Auger South photon conversion
is important for a sizeable part of the sky
(Figure~\ref{fig-pc}, right).
As expected, small conversion probabilities are found for sky regions
around the pointing direction of the local magnetic field vector.
Thus, we expect significant differences in $P_{\rm conv}$
between the two sites for a given direction in terms of local
coordinates, with the difference being strongly dependent on the
regions of the sky. Also shown in Figure~\ref{fig-pc}
(bottom plots) are therefore sky maps of the {\it difference}
in $P_{\rm conv}$
between Auger North and South for the two fixed energies.
While for the larger part of the sky, $P_{\rm conv}$ is
larger at Auger North (due to the stronger field), there
is at 100~EeV a significant part of the sky where
$P_{\rm conv}$ is larger at Auger South (due to the
different pointing of the magnetic field vector).

It is clear from this that cuts on the local shower arrival
direction can be introduced to select regions of the sky 
where $P_{\rm conv}$ is larger (or smaller) at one site
compared to the other site. Applying the same cuts
to data from the two sites, event sets with enriched
(for the one site) and depleted (for the other site) fractions
of converted (or unconverted) photons can be prepared.
A possible photon signal could then show up with
{\it different} signatures at the two sites for the
{\it same} selection cuts. This will be studied in more detail
in Section~\ref{sec-scenarios}.

Another feature is showing up in Figure~\ref{fig-pofce}
where sky maps of energies are plotted for fixed $P_{\rm conv}$.
The range of energies covered (for the same value of $P_{\rm conv}$)
is larger at Auger North than at Auger South.
For instance, the energy at which $P_{\rm conv} = 10\%$ is reached
changes at Auger North by a factor $\sim$15 over the sky, at Auger South
only by a factor  $\sim$4.
In other words, the sky pattern for geomagnetic cascading is more
``inhomogeneous'' at Auger North (larger transition range
in energy from small to large conversion probability) than at Auger South.

In fact, while the preshower effect starts
at smaller energies for a larger part of the sky at Auger North,
the transition range to full conversion extends
to even {\it higher} energies at Auger North,
see Figure~\ref{fig-pofce} (bottom), sky map for 
$P_{\rm conv} = 90\%$.
As an example, a 250~EeV photon ($\lg(E/$eV)=20.4)
has a conversion probability  $>40\%$ for all directions
at Auger South, while at Auger North a small window with
$P_{\rm conv} < 10\%$ still exists (see Figure~\ref{fig-app3}).

At first glance this seems unexpected. As the magnetic field
strength is a factor two
larger at Auger North, one would rather expect photon conversion
at smaller energies at Auger North. However, this feature is again
related to the different locations within the geomagnetic field,
with Auger North being considerably closer to the magnetic pole:
in this case, the local magnetic field lines are less curved when
following them from ground to higher altitudes.
Correspondingly, the {\it transverse} projection of the magnetic
field vector, which determines the local conversion probability,
remains small out to considerably larger distances from the Earth
compared to the situation at Auger South.
Thus, the fact that Auger North is closer to the pole results
in a larger transition region for photon conversion,
starting at smaller energies
compared to Auger South (due to the {\it larger strength}
of the local magnetic field)
and ending at higher energies compared to Auger South
(due to the {\it smaller bending} of the magnetic field lines with
distance). 
%In Appendix~\ref{app-a}, these features are shown to
%be qualitatively expected within a simple model of a
%magnetic dipole.

To summarize, the different positions of the sites in the
geomagnetic field translate into a spatial shift of the
conversion patterns at the sky according to the pointing of
the local magnetic field vector and into modifications of the
transition energy range. The modifications of the transition energies
are not a simple shift but also the size of the energy range 
is affected.

\subsection{Particle energy spectra in converted showers}

\begin{figure}[t]
\begin{center}
\includegraphics[height=8.7cm,angle=0]{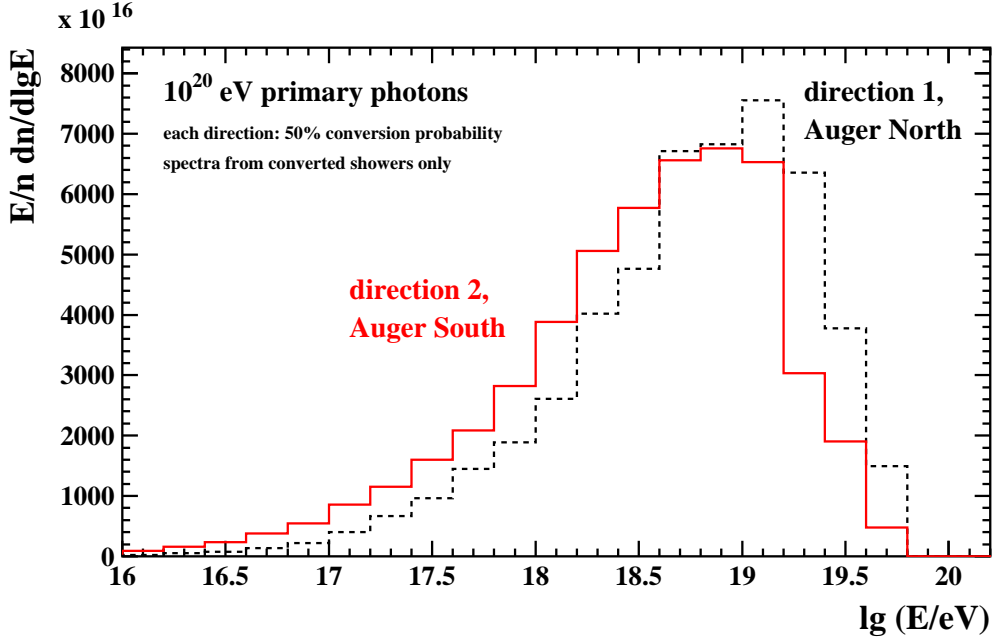}
\caption{
Average energy spectrum of preshower particles, weighted by particle
energy (converted events only) for 10$^{20}$~eV photons.
Chosen are two directions, one at each site, with the same integrated
conversion probability $P_{\rm conv} = 50\%$.
Direction~1 (Auger North): 
$\theta \sim 60^\circ , \phi \sim 270^\circ$,
direction~2 (Auger South):
$\theta \sim 66^\circ , \phi \sim 150^\circ$.
}
\label{fig-espek}
\end{center}
\end{figure}

Here we check to what extent it is sufficient to consider only
the difference in conversion probability when comparing the two sites.
For a given energy, two different
directions with the same integrated conversion probability $P_{\rm conv}$
correspond to different spatial trajectories through the magnetosphere.
One could imagine in one case a relatively small transverse
magnetic field extending over a larger distance, while in the other
case relatively close to the Earth, a stronger magnetic field contributes.
The same integrated conversion probability could be reached, but the
spectrum of preshower particles may differ (for instance, the synchrotron emission
spectrum gets harder with stronger magnetic field).

In Figure~\ref{fig-espek}, averaged spectra of preshower particles at the top
of the atmosphere are shown,
weighted by the particle energy, for different directions at Auger North
and South.
In both cases, $P_{\rm conv} = 50\%$.
One can see that both spectra look quite similar; most energy is carried
by particles of $\sim 10^{19}$~eV. A shift between the distribution of
about a factor 1.6 in energy can be noted, however.
Performing air shower simulations for these two directions, a difference
in the average depth of shower maximum $X_{\rm max}$ of the converted events of 
$\sim 20$~g~cm$^{-2}$ is obtained.
This difference is much below the shift in $X_{\rm max}$ comparing converted
and unconverted photon showers to each other (see next Section).
For other directions of same integrated conversion probability,
differences are typically smaller than shown here.
For our purposes in this work, we conclude
that the differences in conversion probabilities are a sufficient
measure for the different preshower characteristics when comparing the
two sites.

%%%%%%%%%%%%%%%%%%%%%%%%%%%%%%%%%%%%%%%%%%%%%%%%%%%%%%%%%%%%%%%%%%%%%
% 33333333333333333333

\section{Air showers initiated by converted and unconverted photons}
\label{sec-eas}

In this section, we study the basic features of air showers initiated by
converted and unconverted photons, with particular emphasis on the
{\it differences} between these two classes of events. 
The reasoning is that the differences between Auger South and North
in the sky patterns for photon cascading, studied in the preceeding
Section, translate into differences of the air shower features that
are well observable.

It is well known that unconverted showers, contrary to converted,
have a considerably delayed
shower development due to the LPM effect~\cite{lpm}.
Additionally, event-by-event
fluctuations can be extraordindarily large due to a positive correlation
of the suppression of the cross-section with air density (a photon traversing a
certain depth $\Delta X$ without an interaction has a larger probability for
traversing an additional $\Delta X$ due to the larger air density).

\begin{figure}[t]
\begin{center}
\includegraphics[height=8.7cm,angle=0]{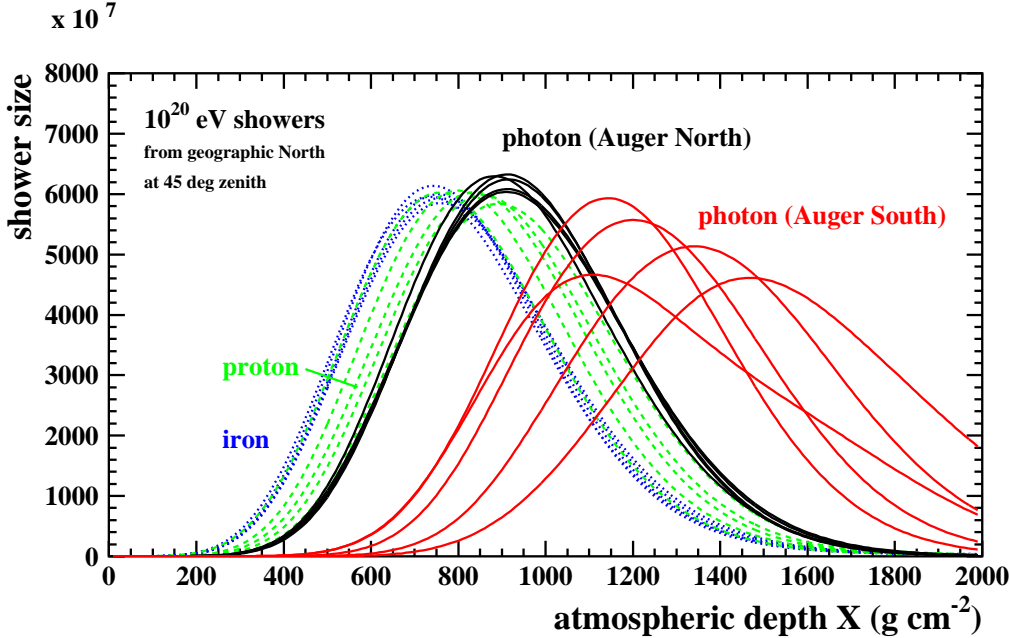}
\caption{
Examples of shower profiles (randomly selected) for $10^{20}$~eV photons at Auger North
(solid black line), at Auger South (solid red line),
protons (dashed green line), and iron nuclei (dotted blue line).
For all events, arrival directions are fixed to local geographic North
at 45$^\circ$ zenith.
}
\label{fig-profiles}
\end{center}
\end{figure}

\begin{figure}[t]
\begin{center}
\includegraphics[height=8.7cm,angle=0]{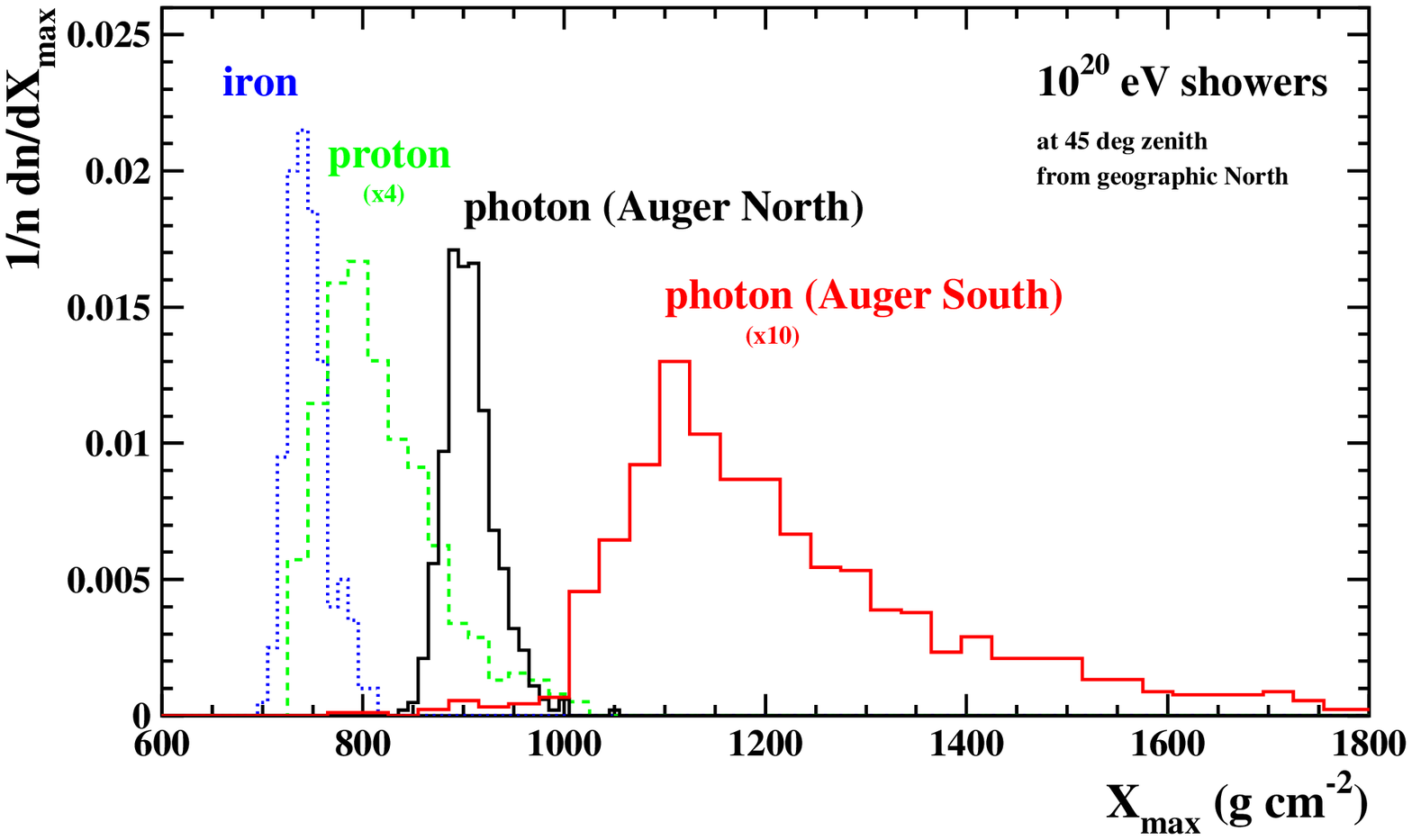}
\includegraphics[height=8.7cm,angle=0]{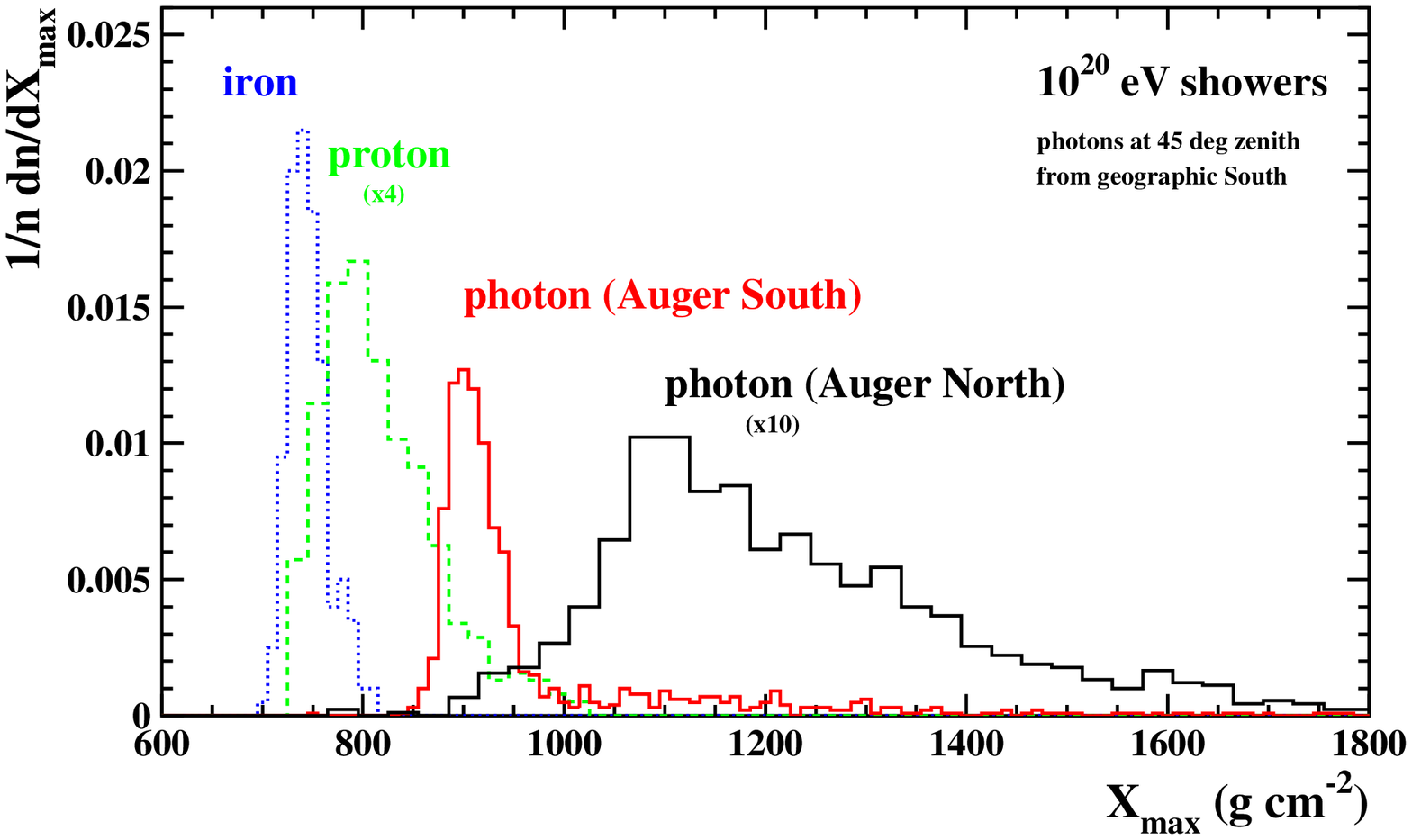}
\caption{
{\it Upper panel}:
$X_{\rm max}$ distributions of different primaries for
the same conditions as in Figure~\ref{fig-profiles}
(photons arriving from local geographic North).
{\it Lower panel}: same as in upper panel, but the photon simulations
were performed with the azimuth changed by 180$^\circ$
(photons arriving from local geographic South).
If indicated, distributions were scaled.
}
\label{fig-xmax}
\end{center}
\end{figure}

In Figure~\ref{fig-profiles}, examples of shower profiles are shown
for $10^{20}$~eV primaries.
Simulations are carried out with CONEX,
which reproduces well CORSIKA~\cite{heck} results. The extrapolation of the photonuclear
cross section is based on the fit by the Particle Data Group~\cite{pdg}.
All primaries arrive from the same local direction, in this
case from geographic North at 45$^\circ$ zenith.
The photon conversion probability is large at Auger North ($>99.9$\%)
and small at Auger South (0.4\%). Consequently, most of the photons
at Auger North convert and have a depth of shower maximum 
200$-$300~g~cm$^{-2}$ smaller than the (mostly unconverted)
photon showers at Auger South. As expected, also fluctuations are
much smaller at Auger North in this example.
The corresponding distribution of depth of shower maximum $X_{\rm max}$
from simulating 1000 events for the conditions
in Figure~\ref{fig-profiles} is shown in Figure~\ref{fig-xmax} (upper panel).
For the same local direction, the expected features of photon showers
are very different at the two sites. 

For comparison, also profiles for hadron primaries
are added using the QGS\-JET~01 model~\cite{qgs01}.
Converted photons are more similar to protons, but still
reaching $X_{\rm max}$ at larger depths.
More relevant for the present study are, however, the differences
between converted and unconverted photons.

According to the results of the preceding Section, we expect the
opposite behavior (larger $X_{\rm max}$ at Auger North) for certain other
directions. Results of simulations with the same conditions
as in Figure~\ref{fig-profiles} and in the upper panel of
Figure~\ref{fig-xmax}, but with the azimuth changed by 180$^\circ$,
are presented in the lower panel of Figure~\ref{fig-xmax}.
The angle between this direction and the
local magnetic field vector at Auger North
is relatively small. As expected,
the $X_{\rm max}$ distribution of photons at Auger South is now peaked at smaller
values, while the distribution at Auger North is dominated by
the large $X_{\rm max}$ values from unconverted events.

\begin{figure}[t]
\begin{center}
\includegraphics[height=8.7cm,angle=0]{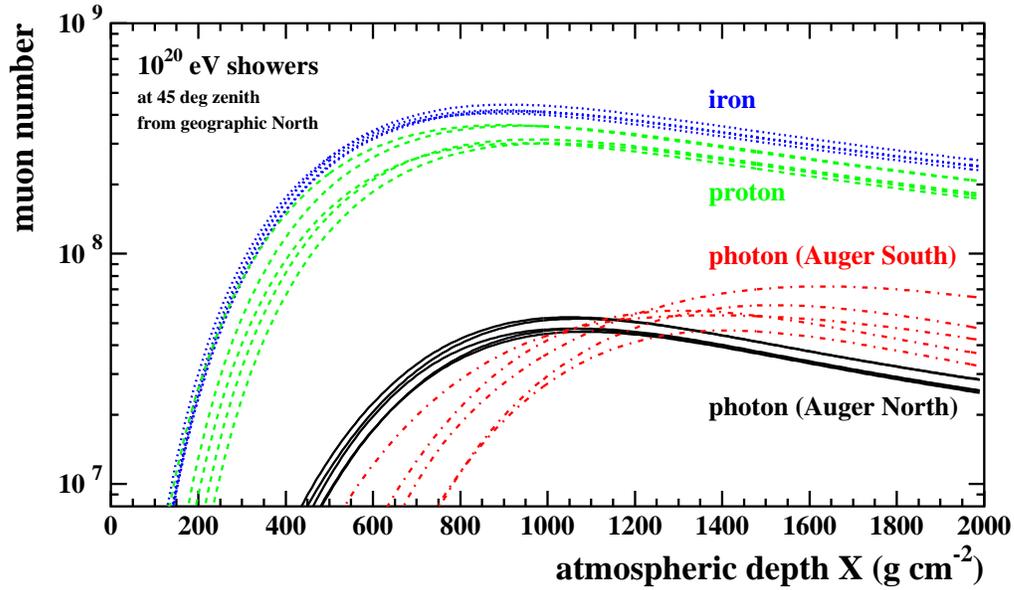}
\caption{
Examples of muon profiles (randomly selected) for $10^{20}$~eV primaries for a muon energy threshold 1 GeV
(same events as shown in Figure~\ref{fig-profiles}).
}
\label{fig-muonprofiles}
\end{center}
\end{figure}

Another important quantity to distinguish photons from hadron primaries
is the muon content of the air shower. It is well known that photon showers
have less muons compared to hadron-initiated ones; important for the
current investigations is whether a significant difference in the
muon content exists between converted and unconverted photons.
In Figure~\ref{fig-muonprofiles}, for the same events shown in
Figure~\ref{fig-profiles}, the longitudinal profiles of muons
are displayed. One can see a shift between the muon profiles for
Auger North (all converted here) and Auger South (all unconverted here).
This shift is, to a large extent, connected to the shift in
$X_{\rm max}$ (Figure~\ref{fig-profiles}).

\begin{figure}[t]
\begin{center}
\includegraphics[height=8.7cm,angle=0]{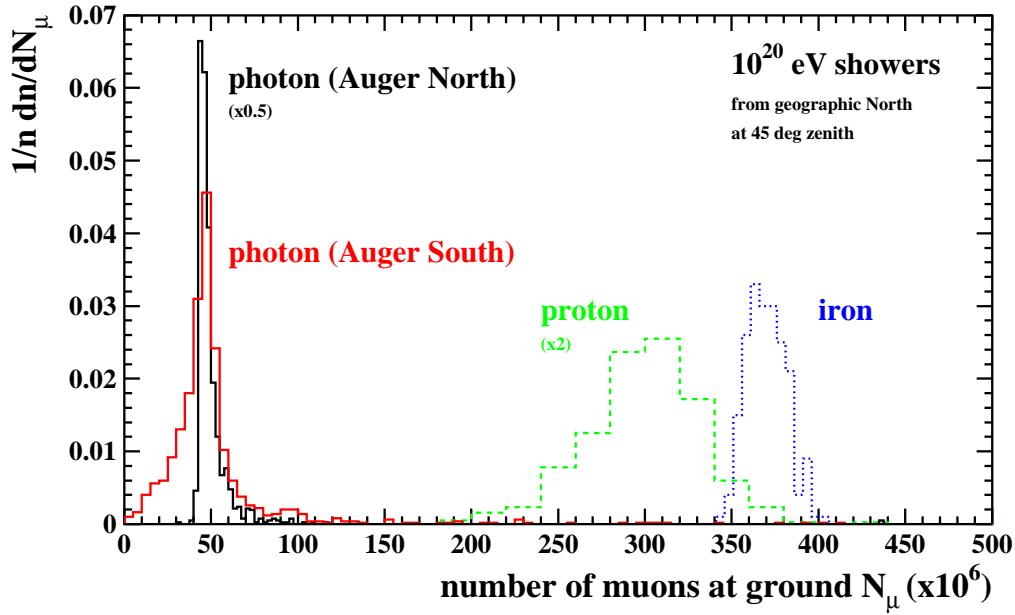}
\caption{
Distributions of total number of muons at ground (energy threshold 1 GeV)
for different primaries for the same conditions as in 
Figures~\ref{fig-profiles} and \ref{fig-xmax} (upper panel) and \ref{fig-muonprofiles}.
If indicated, distributions were scaled.
}
\label{fig-muon}
\end{center}
\end{figure}

The distribution of the total number of muons at ground
is given in Figure~\ref{fig-muon}. As expected, fluctuations are
larger in case of unconverted events, i.e.~at Auger South in this example. 
However, average values
are quite similar. Thus, 
differences between converted and unconverted photon showers
in the number of muons are much less spectacular than in $X_{\rm max}$.
Part of the differences can directly be related to $X_{\rm max}$.
We conclude that the main difference between converted and
unconverted photons, and thus the main difference that could
be exploited by observing from two sites, can be characterized
by the $X_{\rm max}$ of the shower.


There are other observables from ground arrays that were shown
(or are expected) to differ between converted and unconverted
photons, namely
the shower size (total number of electrons at ground)~\cite{stanev97},
the curvature of the shower front and the steepness of the lateral
distribution of ground particles~\cite{bertou00},
or the risetime of the detector signal at a certain distance from
the shower core~\cite{augerphoton}.
The differences between these observables for the two classes of photon
events are closely connected to the stage of shower development
and, correspondingly, to $X_{\rm max}$.
We therefore expect, firstly, that also data from ground arrays
alone, even without a direct observation of $X_{\rm max}$ such as from
fluorescence telescopes, can be exploited to search for
complementary photon signatures at the two sites.
And secondly, a study of $X_{\rm max}$ distributions (which provides a measure
of the relative shower age at ground)
provides us with
most relevant information for investigating possible complementary
features of both sites.

%%%%%%%%%%%%%%%%%%%%%%%%%%%%%%%%%%%%%%%%%%%%%%%%%%%%%%%%%%%%%%%%%%%%%
% 44444444444444444444

\section{UHE photon scenarios and their observation at Auger South and North}
\label{sec-scenarios}

We now give an illustration of how advantage could be taken of the
complementarity between the preshower characteristics at Auger North
and South.
As examples, we consider a diffuse photon signal, a signal from a
source region,
and the absence of photons (i.e.~determining upper limits).

\subsection{Diffuse photon signal}

%In this example, we assume an isotropic primary flux with the
%all-particle energy spectrum according to the first estimate of the
%Auger South Observatory~\cite{sommers}.
%More specifically, we take a
%power law with spectral index -2.84 which we extrapolate to highest energies. 
%ph 06.10.04
%Such a spectrum is consistent with the first Auger data.
%%%%%%%%%%%%%%%%
In this example, we assume an isotropic primary flux with the
all-particle energy power low spectrum with index -2.84. Such a spectrum is
consistent with the first estimate from the Auger South Observatory~\cite{sommers}.
It should be noted that a deviation from this power law, for instance a
suppression of the flux above the energy of the speculated GZK cut-off,
would result in a correspondingly modified all-particle flux at
highest energies.
%ph 06.10.04
The existence of such a hypothetic suppression will not affect 
the conclusions of this section: discussed differences between photon showers 
observed at Auger North and South depend only on the local geomagnetic 
field (see below for a quantitative example of applying a simple cut-off to the energy spectrum). 
%%%%%%%%%%%%%%%%%%%%

We assume protons and photons as primaries. 
The input fraction of photons as a function of primary energy 
follows the results from a topological defect model in~\cite{models}.
More specifically, the photon fraction is 9\%, 15\%, 27\%, and 48\%
for energies above $\lg (E/$eV) = 19.0, 19.3, 19.6, and 20.0,
respectively (see also the curve labeled ``TD'' in Figure 8
in Ref.~\cite{augerphoton}).

For each Auger site, we simulated $\sim$1000 events above
$\lg (E/$eV) = 19.6 with zenith angles between 30$-$75$^\circ$ and random azimuth.
We accounted for a detector resolution of 25~g~cm$^{-2}$
in $X_{\rm max}$ and 10\% in primary energy.
The zenith angle range was chosen similar to that in the analysis
of Auger data for the first photon limit~\cite{augerphoton}.
In particular,
near-vertical events are excluded in this example to reduce
the fraction of photon events with $X_{\rm max}$ below ground
(see also discussion in Section~\ref{sec-absofphot}).
The simulated event number corresponds to an observation time
of about 5$-$8 years of the completed Auger South ground array.
For a similar data taking period, data statistics of the
fluorescence detectors is reduced by a factor $\sim$7.

\begin{figure}[t]
\begin{center}
\includegraphics[height=8.7cm,angle=0]{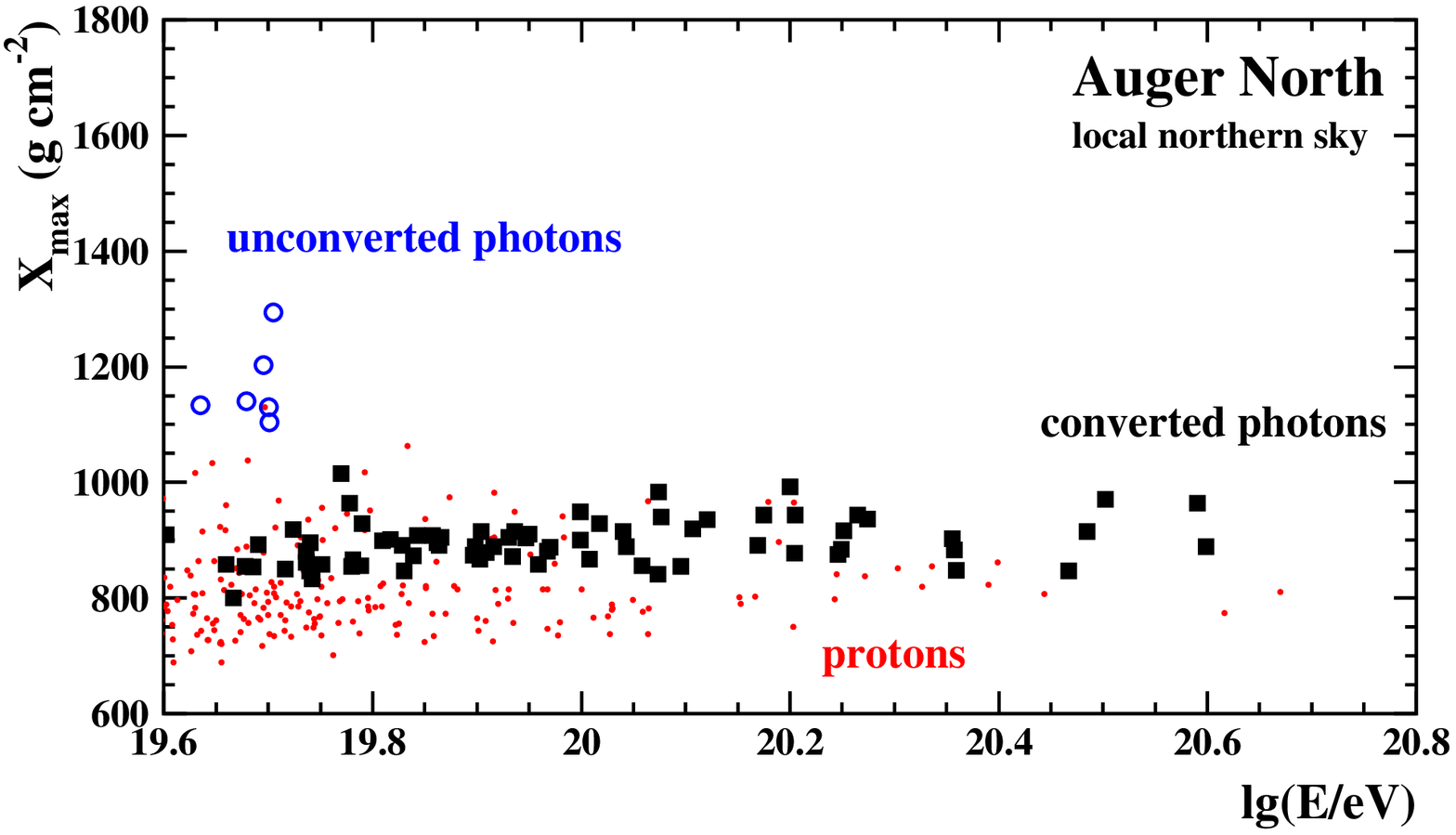}
\includegraphics[height=8.7cm,angle=0]{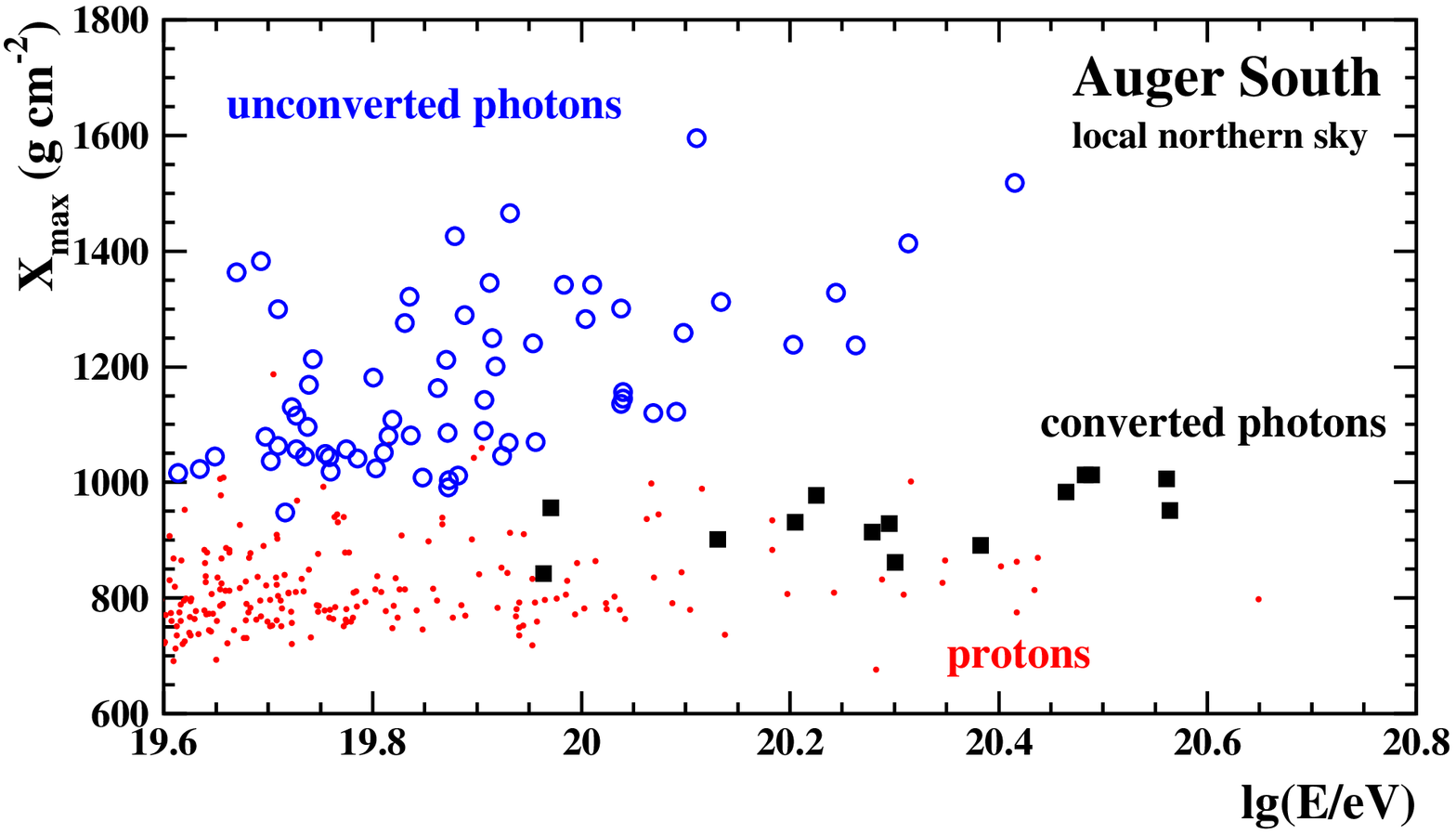}
\caption{
$X_{\rm max}$ versus energy for events arriving from the local
{\it northern} sky (see text for definition)
at Auger North (top) and at Auger South (bottom),
for unconverted photons (open blue circles), converted photons
(filled black squares), and protons (red dots).
}
\label{fig-xmaxvseln}
\end{center}
\end{figure}

In Figure~\ref{fig-xmaxvseln} we show, as a scatter plot,
the $X_{\rm max}$ versus primary energy. We restricted the
azimuth range in this plot to the local {\it northern} sky
by requiring an azimuth between 30$-$150$^\circ$.
In this region of the sky, photon conversion starts at Auger
North at smaller energies than at Auger South.
As can be seen from the Figure, there are considerably less
events with large $X_{\rm max}$ (e.g.~exceeding 1000~g~cm$^{-2}$)
at Auger North, for the same overall cuts applied to the
data at both sites.

\begin{figure}[t]
\begin{center}
\includegraphics[height=14.0cm,angle=0]{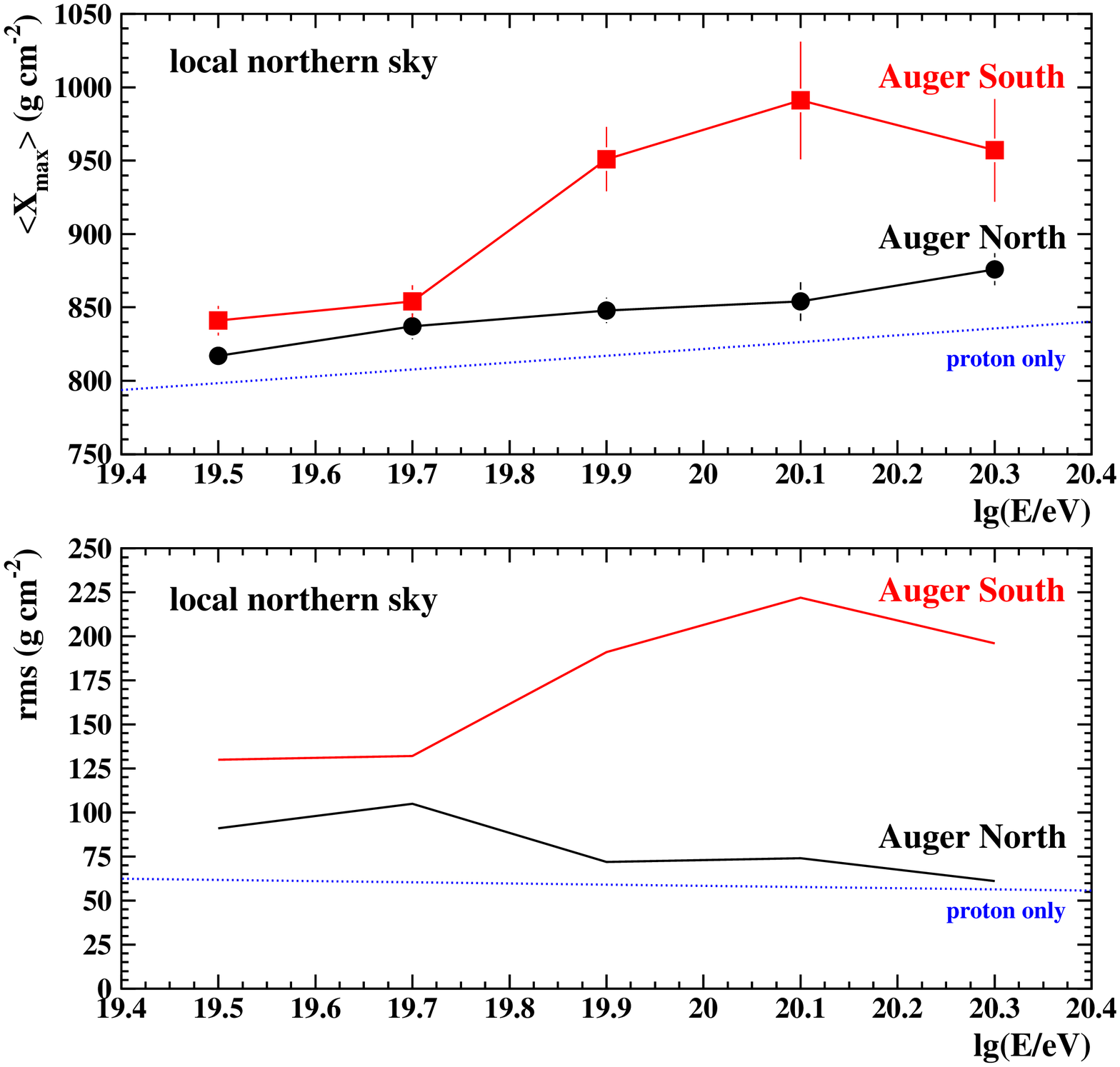}
\caption{
Average $X_{\rm max}$ (top) and rms of $X_{\rm max}$ (bottom)
versus energy for the conditions shown in Figure~\ref{fig-xmaxvseln},
i.e.~local northern sky at Auger North and at Auger South.
For comparison, values corresponding to a pure proton flux are
also shown using the model QGS\-JET~01 (dotted blue lines).
}
\label{fig-avxmaxvseln}
\end{center}
\end{figure}

The corresponding average $X_{\rm max}$ and the $X_{\rm max}$ fluctuations
as a function of energy are shown in Figure~\ref{fig-avxmaxvseln}.
Clear differences between the two sites can be noted. 
In particular, the larger average $X_{\rm max}$ at Auger South
is accompanied by significantly increased shower fluctuations.

\begin{figure}[t]
\begin{center}
\includegraphics[height=8.7cm,angle=0]{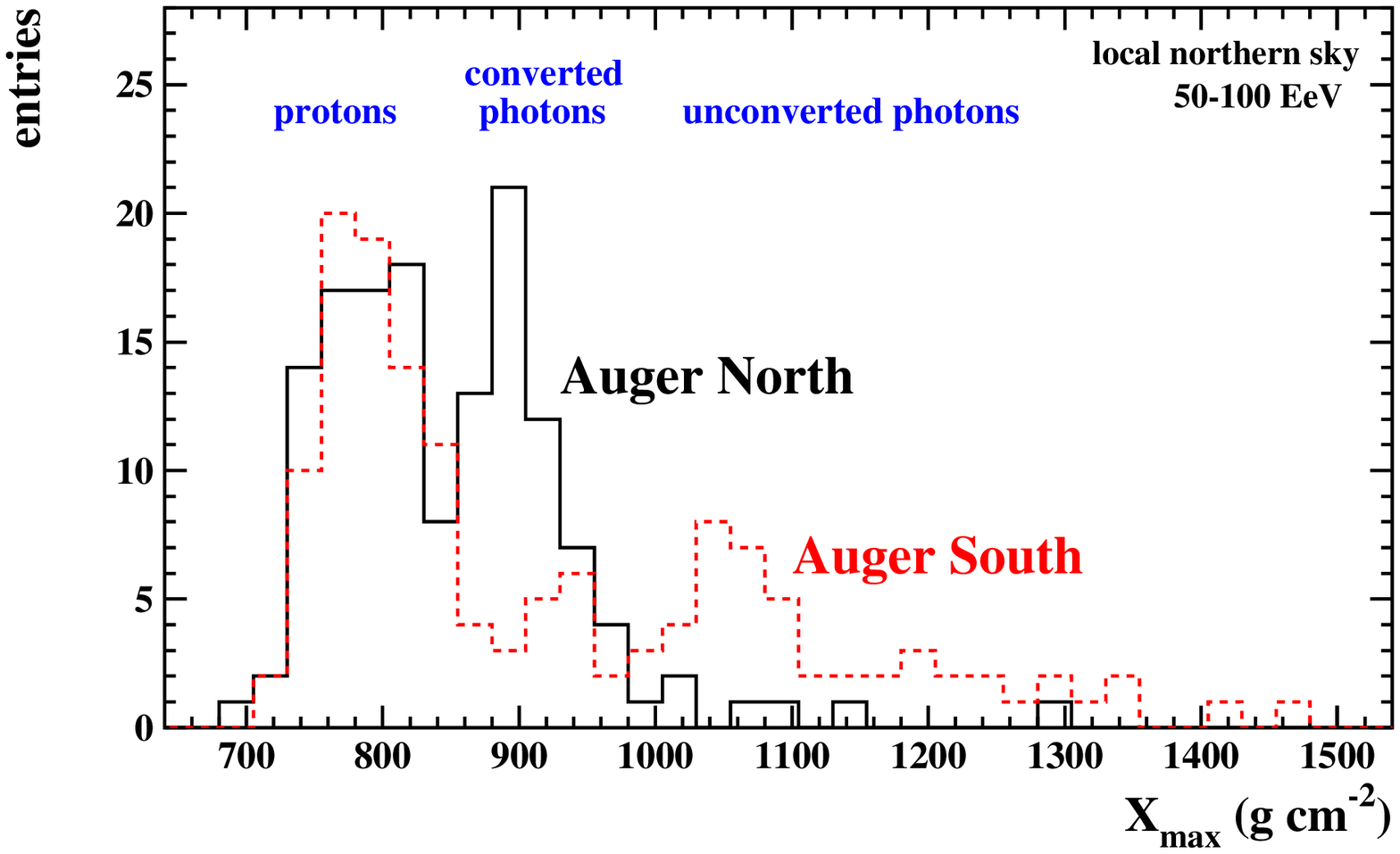}
\caption{
$X_{\rm max}$ distributions for Auger North and South
at the local northern sky at 50$-$100~EeV
(projection from the scatter plot shown in Figure~\ref{fig-xmaxvseln}).
The different primary types (protons, converted and unconverted
photons) contribute mainly to ranges in $X_{\rm max}$ as
indicated in the plot. In particular, the peak in the distributions
at $X_{\rm max} \sim 800$~g~cm$^{-2}$ comes from protons,
at $\sim 900$~g~cm$^{-2}$ (at Auger North) from
converted photons, and at $\sim 1050$~g~cm$^{-2}$ (at Auger South) from
unconverted photons.
}
\label{fig-xmaxdistln}
\end{center}
\end{figure}

The $X_{\rm max}$ distribution of events
(as projected from the scatter plot shown in Figure~\ref{fig-xmaxvseln})
is given for 50$-$100~EeV primary energies in Figure~\ref{fig-xmaxdistln}.
At Auger North, the events with large $X_{\rm max}$ (unconverted photons,
present at Auger South), are mostly lacking.
Instead, events accumulate at $\sim 900$~g~cm$^{-2}$ (converted photons).
%ph 06.10.05
To see the effect of a hypothetic GZK suppression of the energy spectrum at high energies,
we applied a simple cut-off at E = 80~EeV ($\lg (E/$eV)~=~19.9) to the data set.
As expected, all the characteristic peaks of the $X_{\rm max}$
distribution remain at the same positions as in case of spectrum {\it without} cut-off
plotted in Figure~\ref{fig-xmaxdistln}.
The only difference is the heights of the peaks corresponding to converted and unconverted photons
with respect to the proton peak: 
they decrease by 10-20\% in case of spectrum {\it with} cut-off. 
This is due to the larger fraction of photons at higher energies
in the assumed model.
%%%%%%%%%%%%%%%%%%%%%%%%%%%%%%%%%%%%%%%%

An observation of such different characteristics at Auger North and South
(signal at different values of $X_{\rm max}$, 
behaviour of average $X_{\rm max}$ and $X_{\rm max}$ fluctuations
as a function of energy),
would be an unambiguous confirmation of a photon signal detection.

So far, the local northern sky was studied. As expected,
a similar but {\it opposite} behaviour is found for the local
southern sky (not shown) when selecting, for instance, shower arrival
directions from azimuth angles between 210$-$330$^\circ$.
Since this occurs at a somewhat higher energy, the event statistics
is reduced in this case due to the steep primary flux spectrum.

\subsection{Photon signal from a localized source region}
\label{subsec-localized-source}

%Let's now regard the case of a photon signal coming from a
%certain source region.
%This is a hypothetical scenario, although it is worthwhile to mention
%that TeV photons were observed by H.E.S.S.~with a relatively hard
%spectrum from sources such as the galactic center region
%(for a recent summary of observations, see for instance~\cite{kappes}).

Let's now regard a hypothetical scenario of a photon signal coming from a
certain source region. Localized hadron-accelerating sources are expected
to produce photons at an energy typically 10 times smaller than the maximum
energy of accelerated hadrons. Detection of the currently highest energy 
event at $3\times10^{20}$~eV provides a motivation for the investigation
of photons at not much lower energies produced by a localized source.

The sky region accessible to both Auger sites is limited to source
declinations $|\delta| <  \sim 53^\circ$. Requiring a maximum zenith angle
of $75^\circ$ for shower observations at both sites, the range of
accessible declinations is $|\delta| <  \sim 38^\circ$.

\begin{figure}[t]
\begin{center}
\includegraphics[height=8.0cm,angle=0]{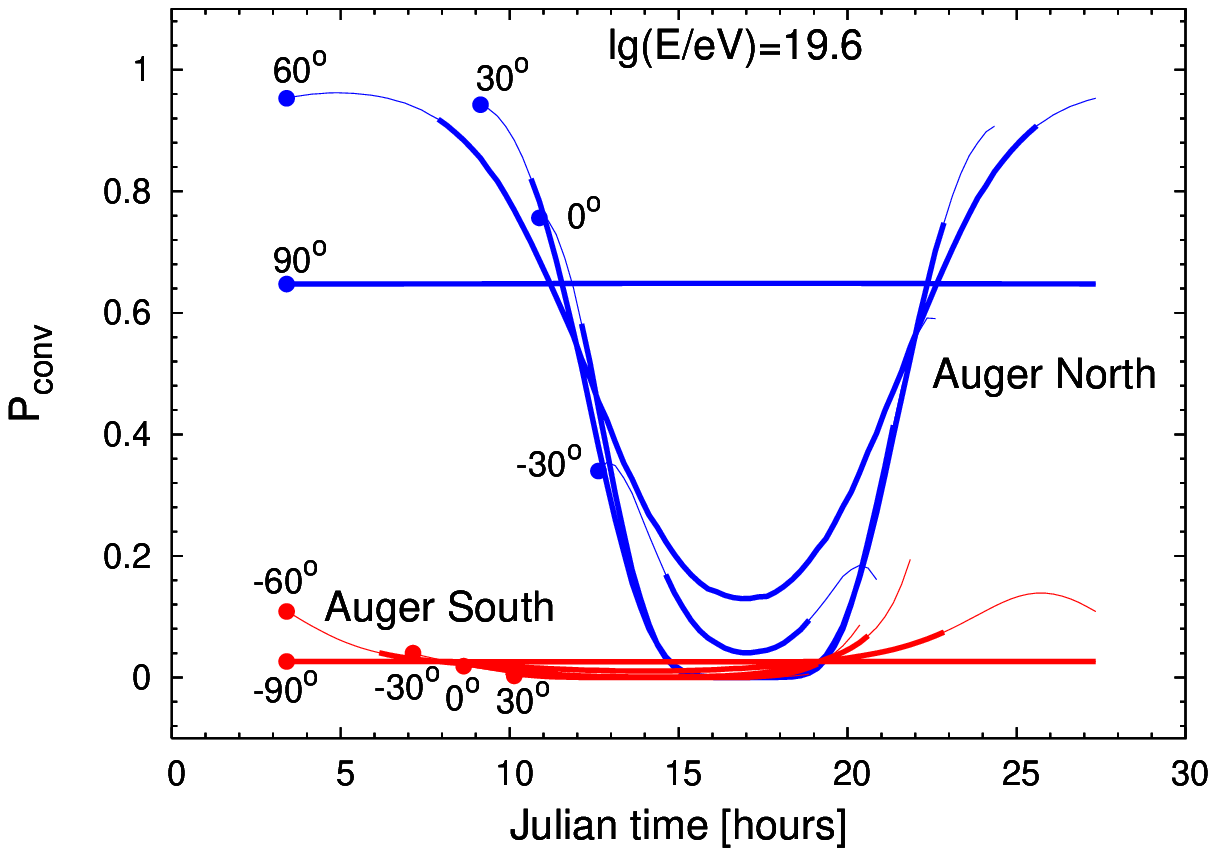}
\includegraphics[height=8.0cm,angle=0]{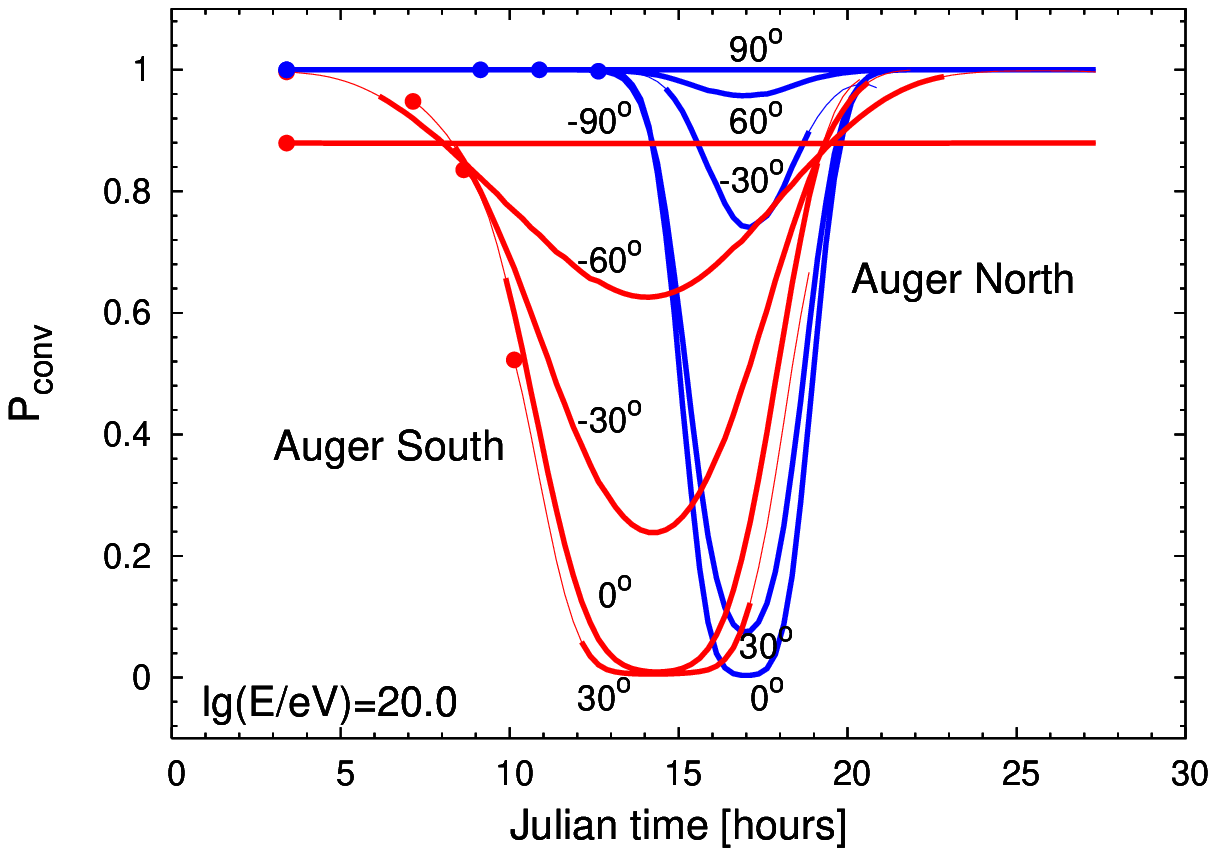}
\caption{
Conversion probability at the two Auger sites versus time (in hours, offset arbitrary)
for $10^{19.6}$~eV (top) and $10^{20.0}$~eV (bottom)
photons for different source declinations as indicated. The same right ascension is assumed for 
all sources. The heavy lines denote the time intervals during which the sources are visible at 
zenith angles less than $75^\circ$. For better visibility the beginnings of lines are marked with large dots.
}
\label{fig-decl}
\end{center}
\end{figure}

The conversion probability of UHE photons from the direction of 
a certain source can change with time even at one site only,
depending on the source declination.
Examples are given in Figure~\ref{fig-decl} for
$10^{19.6}$~eV and $10^{20.0}$~eV photons.
For instance, for a source at the celestial equator 
conversion probabilities between 0$-$75\%
are covered at $10^{19.6}$~eV at Auger North.

It is interesting to note that for the same conditions, the conversion
probability is negligible at Auger South.
Thus, characteristic differences when observing events from the same
sky region exist that again could be used for
a complementary study of UHE photons.
A special case is the scenario of a source region observed
{\it at the same time} from Auger North and South.
Also here, differences exist in the
{\it instantaneous} conversion probability of photons.
This can be seen in Figure~\ref{fig-decl}
by comparing the curves at Auger North and South for the same
source declination as a function of time.
For instance, the instantaneous difference in the conversion
probabilities for $10^{20.0}$~eV photons from a source at the celestial
equator can reach, in the extreme case, almost 100\%.
Differences exist also for sources at declinations $\delta \sim -30^\circ$,
i.e.~for the galactic center region
(Figure~\ref{fig-decl}).

\subsection{Absence of photons}
\label{sec-absofphot}

If the photon flux is below the detection limits of current experiments,
upper limits are deduced by comparing the data to expectations for 
showers initiated by primary photons.
The signatures from photon showers differ between Auger North and South
due to geomagnetic cascading.
Thus, even in case of {\it observing no photons} at both sites,
differences arise in the {\it analysis} of the data
since the photon hypothesis tested when deriving a limit
is different at each site.
Upper limits obtained at Auger South can, to some extend,
be cross-checked this way at Auger North.

In photon limit studies, simulations are used to optimize the analysis
cuts and to calculate the acceptance to photons.
Due to the generally larger
$X_{\rm max}$, the acceptance to photons can be smaller than that to nuclear
primaries when requiring the $X_{\rm max}$ to be above ground.
Then, an efficiency correction to the photon limit may be needed.
To keep the correction small, a fiducial volume cut was applied
in~\cite{augerphoton}, requiring a minimum zenith angle of
35-42 deg (depending on energy).

This condition could be relaxed at Auger North due to the onset
of the preshower effect at smaller energies.
On the one hand, the power of shower-to-shower discrimination
is reduced for events from these arrival directions,
as converted photon showers are more similar to
proton-induced showers than unconverted ones.
On the other hand, the result is a gain in instantaneous aperture
at Auger North.
The effective gain is largest at
energies around $\lg (E/$eV)$\sim$19.8 where for a sizeable part of
the sky, photons already can convert at Auger North, contrary to Auger South.
For identical experiments and requiring $X_{\rm max}$ to be above ground,
the aperture is increased at Auger North by $\sim$30\% (20\%) when
allowing zenith angles up to 60$^\circ$ (75$^\circ$) at $\lg (E/$eV)$\sim$19.8.

%%%%%%%%%%%%%%%%%%%%%%%%%%%%%%%%%%%%%%%%%%%%%%%%%%%%%%%%%%%%%%%%%%%%%
% 55555555555555555555

\section{Conclusion}
\label{sec-conclusion}
The difference of conversion probabilities is a convenient and effective
parameter to estimate the complementarity between two sites for
searching for UHE photons.
Regarding the two sites of the Pierre Auger Observatory, Auger North and
Auger South, significant differences in the preshower features of UHE
photons exist.
The sky patterns of the preshower features are shifted due to the
different pointing directions of the local magnetic field vectors.
In addition, Auger North is located closer to the magnetic pole.
This leads to a larger transition range in energy from small ($<10\%$)
to large ($>90\%$) conversion probabilities compared to Auger South.
More specifically, photon conversion starts at Auger North
at smaller energies due to the larger magnetic field, but conversion probabilities $>90\%$ are
reached for the whole sky at higher energies due to the smaller bending
of the field line with distance.

These differences in the preshower characteristics result in different
rates of (un-)converted photons from the same
(both in local and astronomical coordinates) regions of the sky.
Air showers initiated by converted and unconverted photons
can be well distinguished by current experiments. The main difference
is related to the position of depth of shower maximum
$X_{\rm max}$, which is typically $\sim$200$-$300~g~cm$^{-2}$ smaller
for converted photons.

For a variety of UHE photon flux scenarios (diffuse photon flux;
photons from source regions; absence of photons), the
different preshower characteristics at the experimental sites can
be used for a complementary search for UHE photons.
Most important, a possible detection of UHE photons at Auger South
may be confirmed in an unambiguous way at Auger North by observing
the well predictable change in the signal from UHE photon showers.

%%%%%%%%%%%%%%%%%%%%%%%%%%%%%%%%%%%%%%%%%%%%%%%%%%%%%%%%%%%%%%%%%%%%%

{\it Acknowledgements:}
This work was partially supported by the Polish State Committee for
Scientific Research under grants No.~PBZ~KBN~054/P03/2001 and 
2P03B~11024, by Polish Ministry of Science and Higher Education
under grant N202 090 31/0623 and in Germany by the DAAD under grant No.~PPP~323.
Helpful remarks of the unknown referee, especially those concerning the
motivation of Subsection~\ref{subsec-localized-source}, are kindly acknowledged.
MR acknowledges support from the Alexander von Humboldt foundation.

%%%%%%%%%%%%%%%%%%%%%%%%%%%%%%%%%%%%%%%%%%%%%%%%%%%%%%%%%%%%%%%%%%%%%

\appendix
\section{Transverse component of the geomagnetic field along various trajectories}
\label{app-a}
\begin{figure}[ht]
\begin{center}
% scale temporarily changed from 1.0 to 0.5 MAR0720
\includegraphics[scale=1.0]{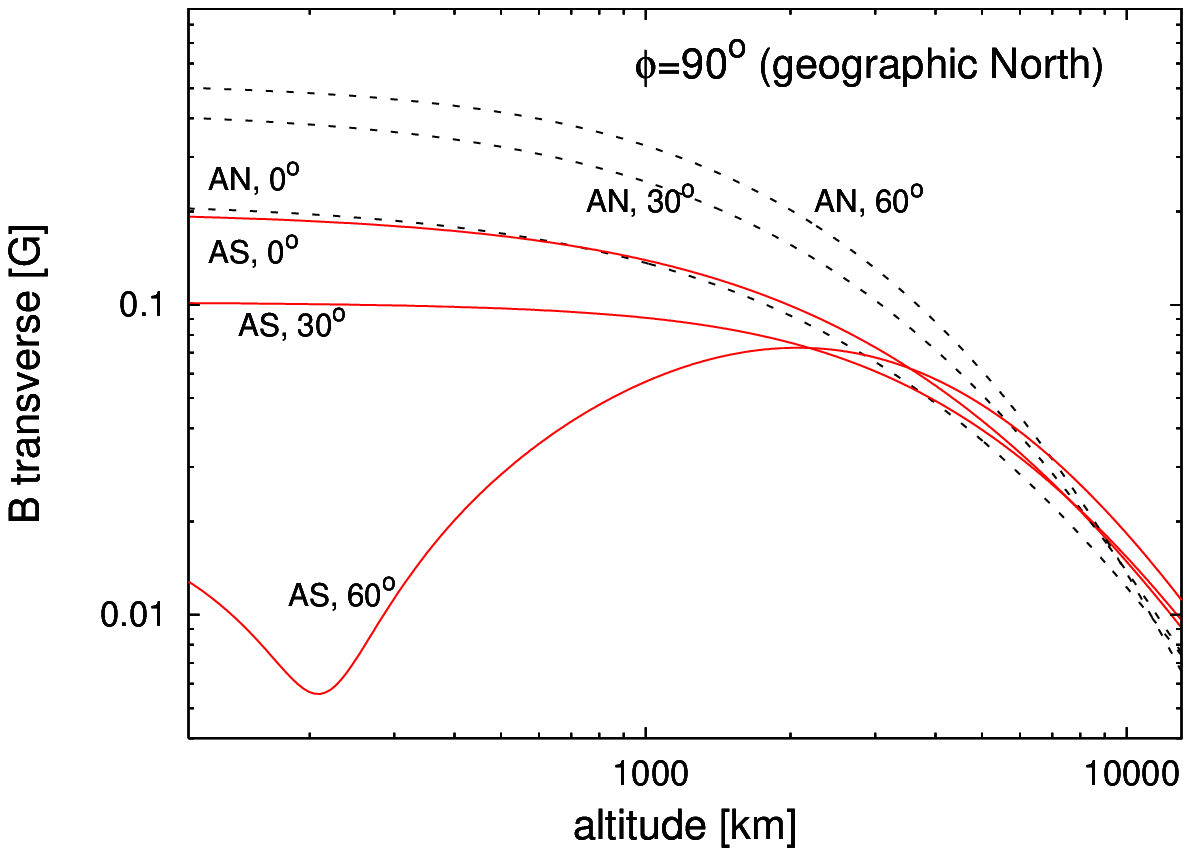}
\includegraphics[scale=1.0]{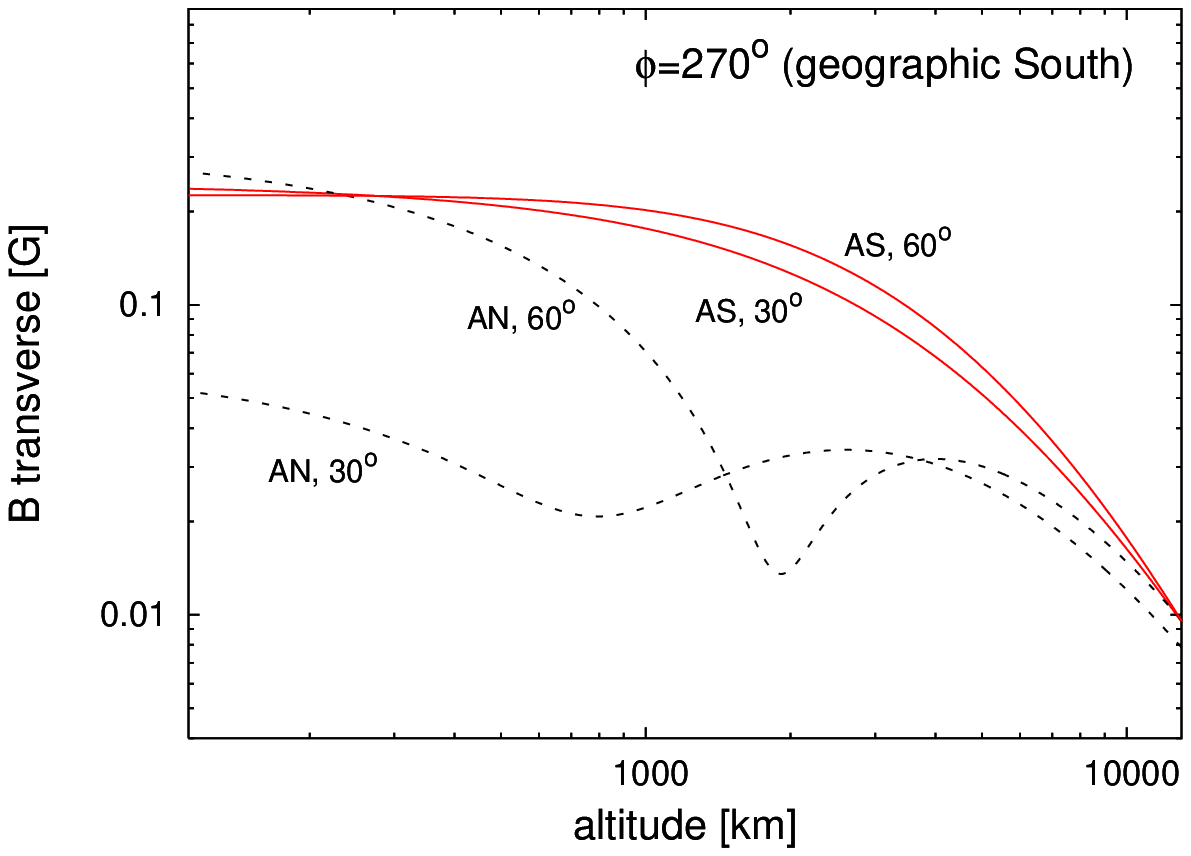}
\caption{\label{bt} Transverse component of the geomagnetic field (according to the IGRF model)
along various directions at two observation sites: Auger South (AS) and Auger North (AN).
The directions are specified by the zenith (attached to each curve)
and azimuth (the same within one plot) angles given in the local coordinate system. The local minima
occur at the altitudes at which the geomagnetic field
is nearly parallel to the considered directions.   
}
\end{center}
\end{figure}

\section{Preshower sky maps}
\label{app-b}

\begin{figure}[h]
\begin{center}
\includegraphics[height=4.8cm,angle=0]{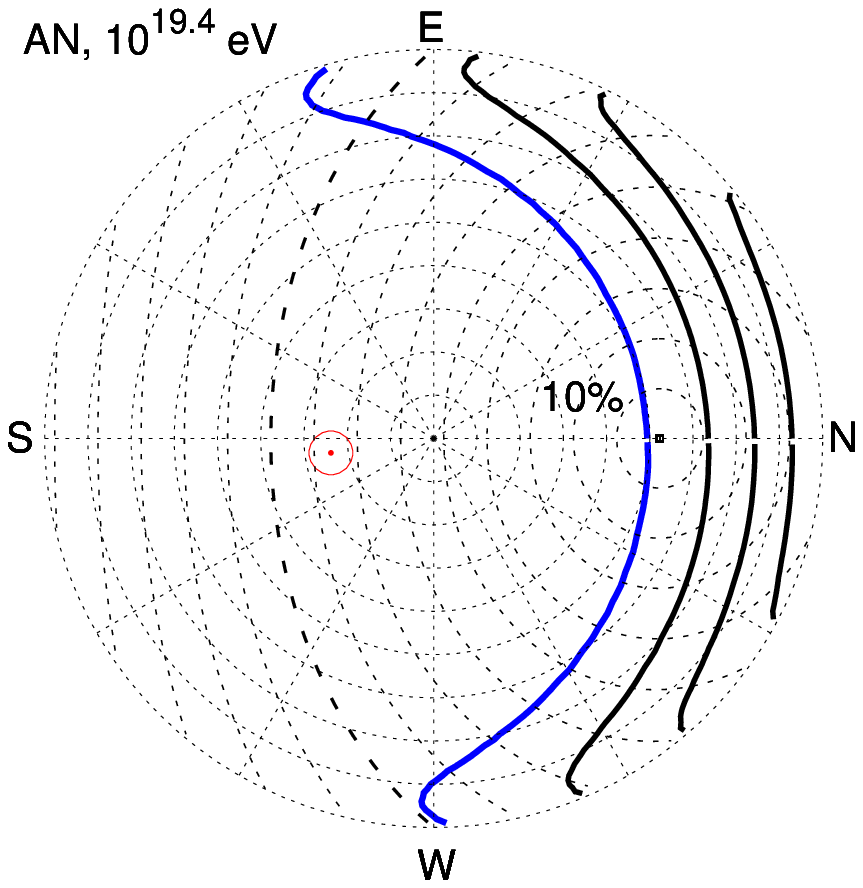}
\includegraphics[height=4.8cm,angle=0]{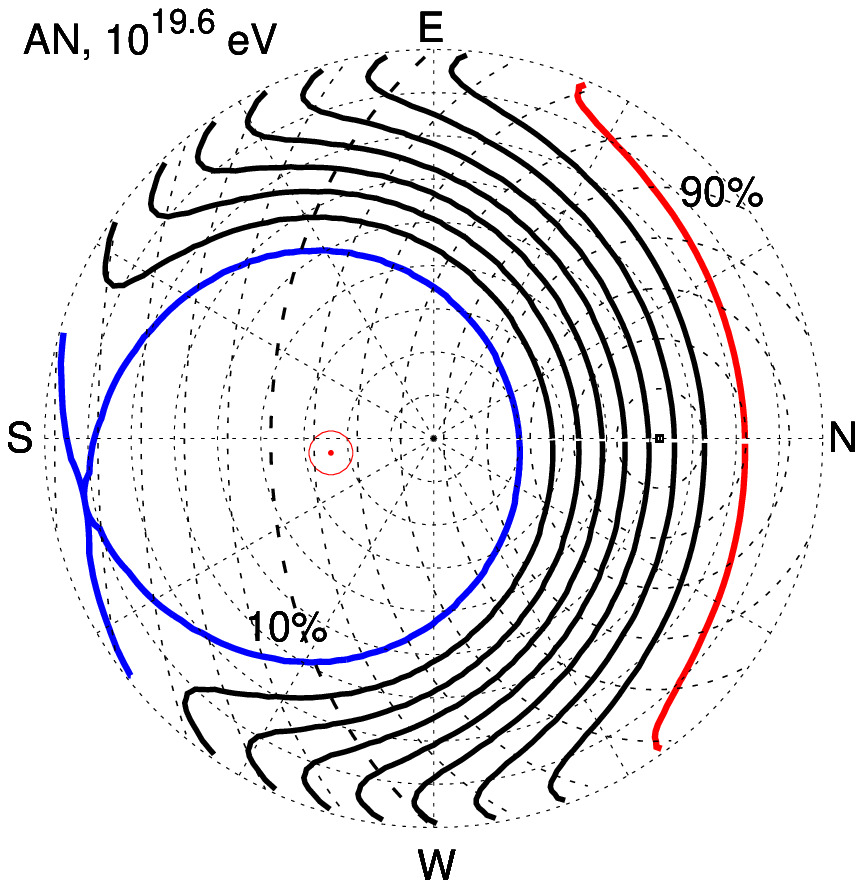}
\includegraphics[height=4.8cm,angle=0]{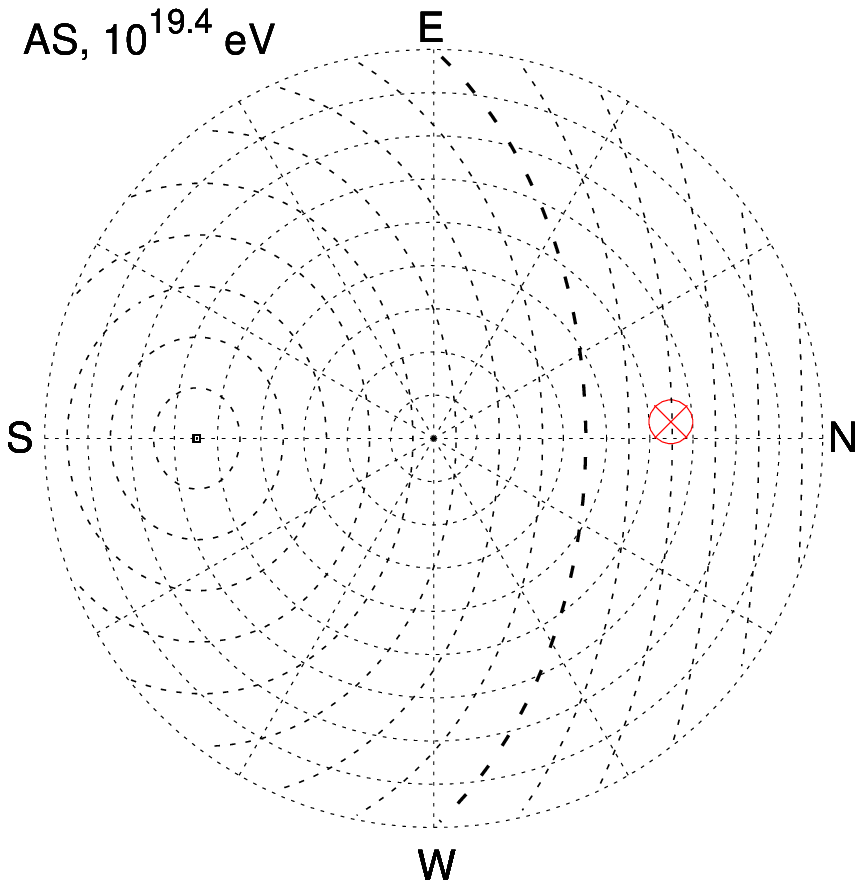}
\includegraphics[height=4.8cm,angle=0]{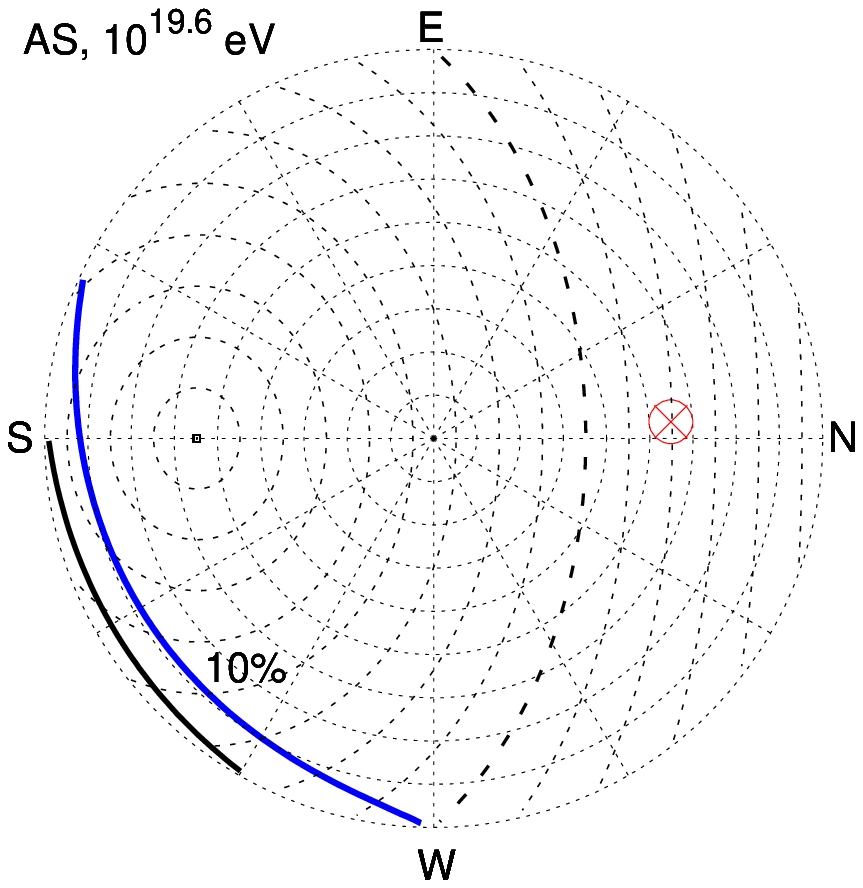}
\includegraphics[height=4.8cm,angle=0]{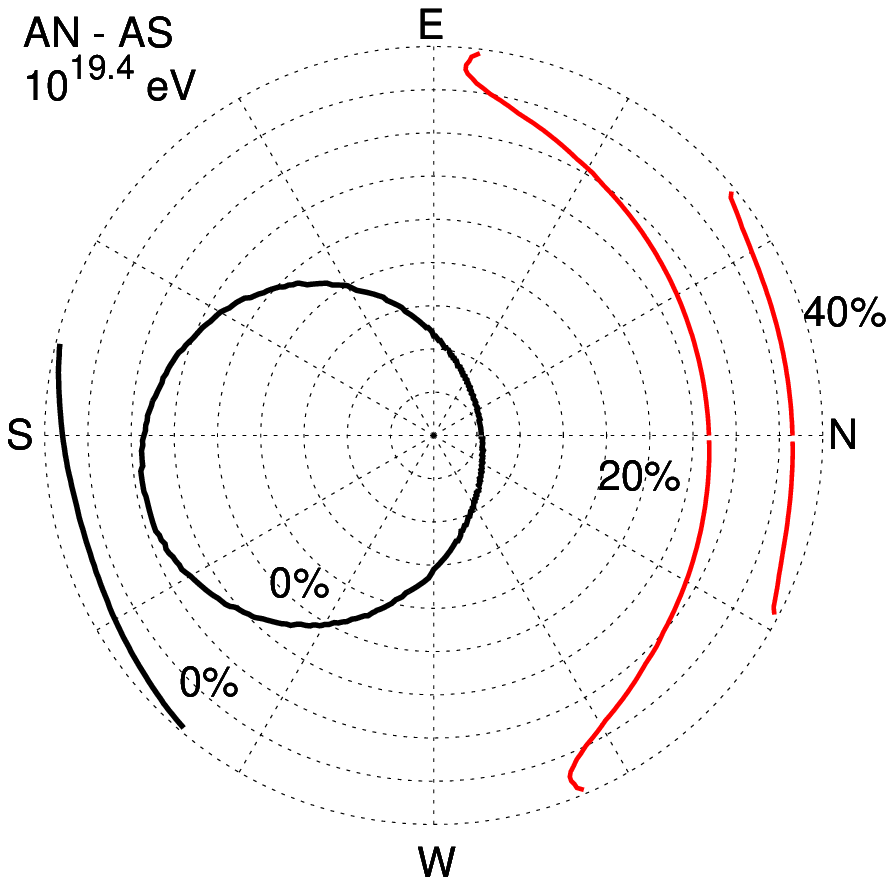}
\includegraphics[height=4.8cm,angle=0]{a_conv_diff_19.6.eps}
\caption{
Sky maps of conversion probabilities at $\lg(E/$eV)=19.4 (left)
and 19.6 (right) for Auger North (AN, top), Auger South (AS, middle), and
the difference (bottom); see also Figure~\ref{fig-pc} for
further explanations.
Additionally shown as dotted black lines are the declinations as
projected on the sky for a given site (top and middle rows of plots)
with a stepsize of 10$^\circ$,
with the celestial equator (declination = 0) highlighted
as a dashed line and the pole indicated as a dot.
}
\label{fig-app1}
\end{center}
\end{figure}
\begin{figure}[h]
\begin{center}
\includegraphics[height=4.8cm,angle=0]{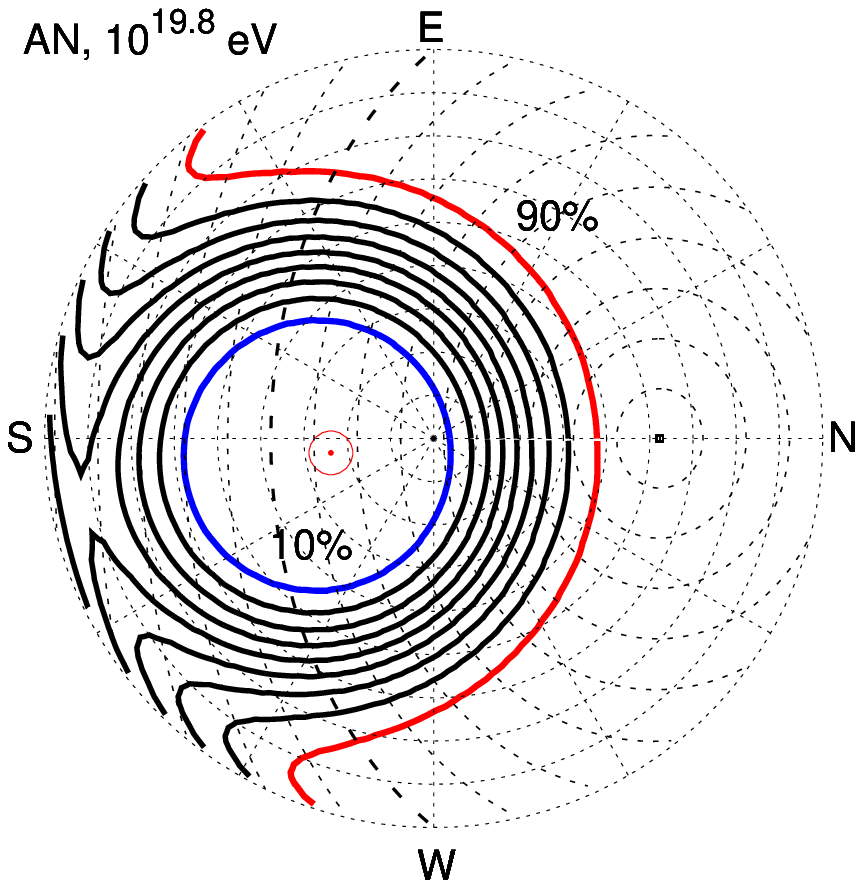}
\includegraphics[height=4.8cm,angle=0]{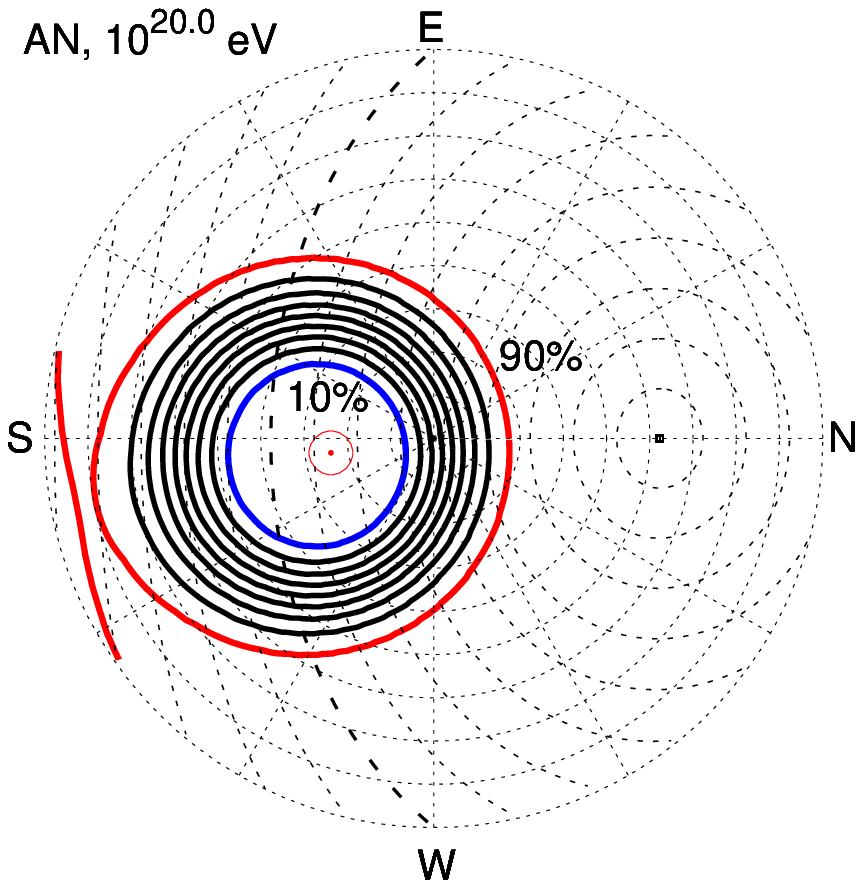}
\includegraphics[height=4.8cm,angle=0]{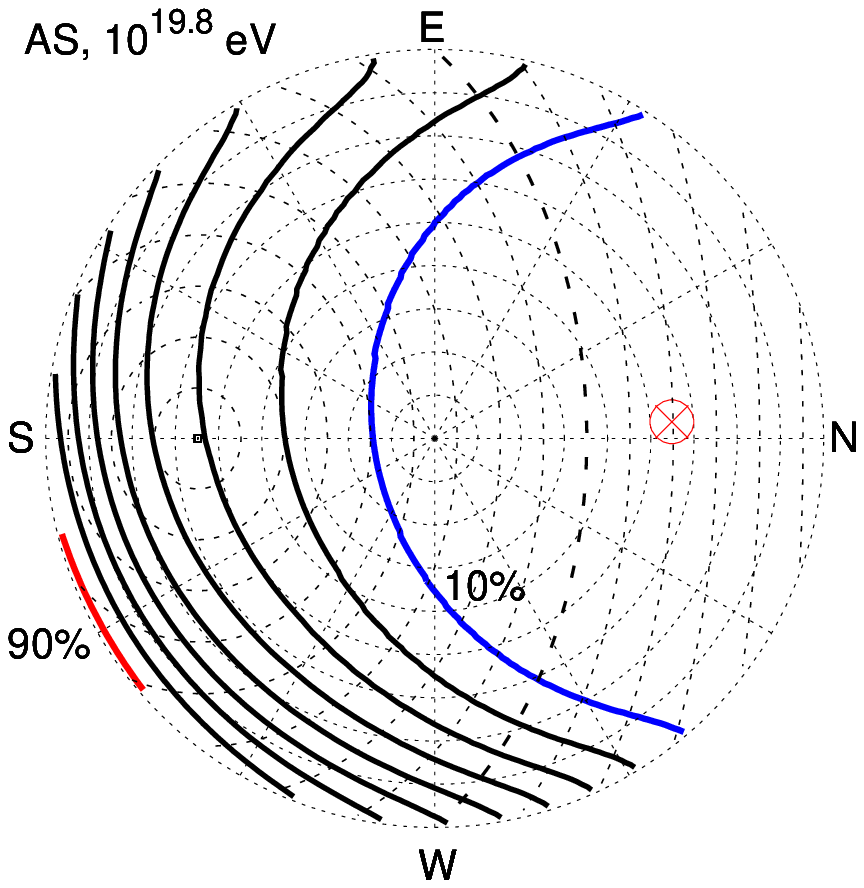}
\includegraphics[height=4.8cm,angle=0]{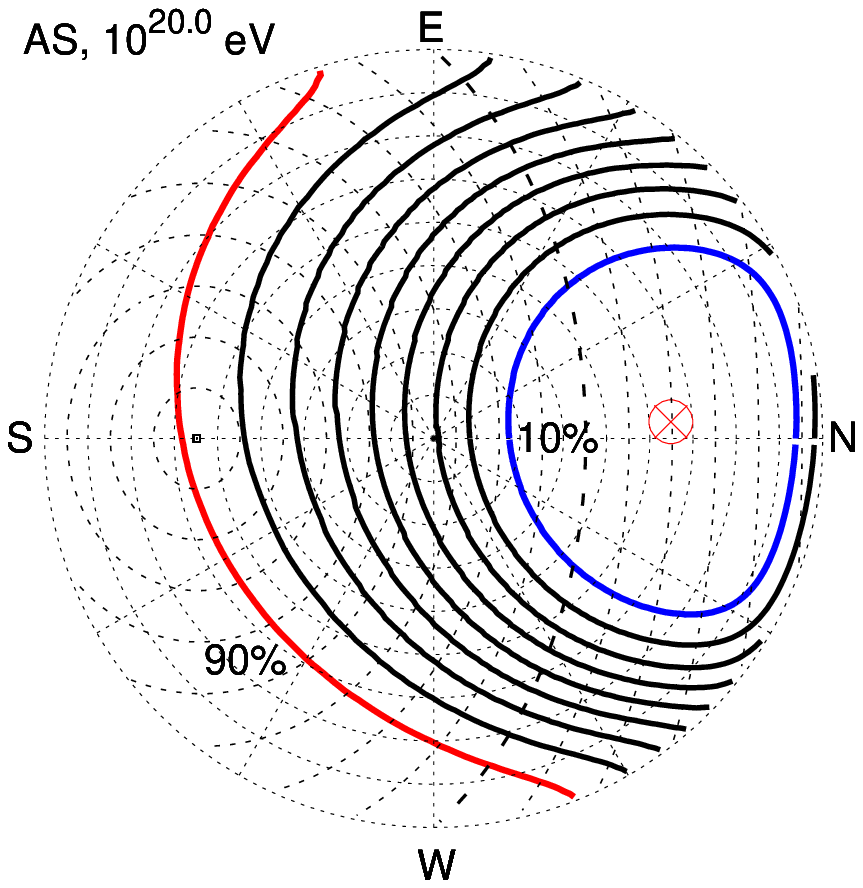}
\includegraphics[height=4.8cm,angle=0]{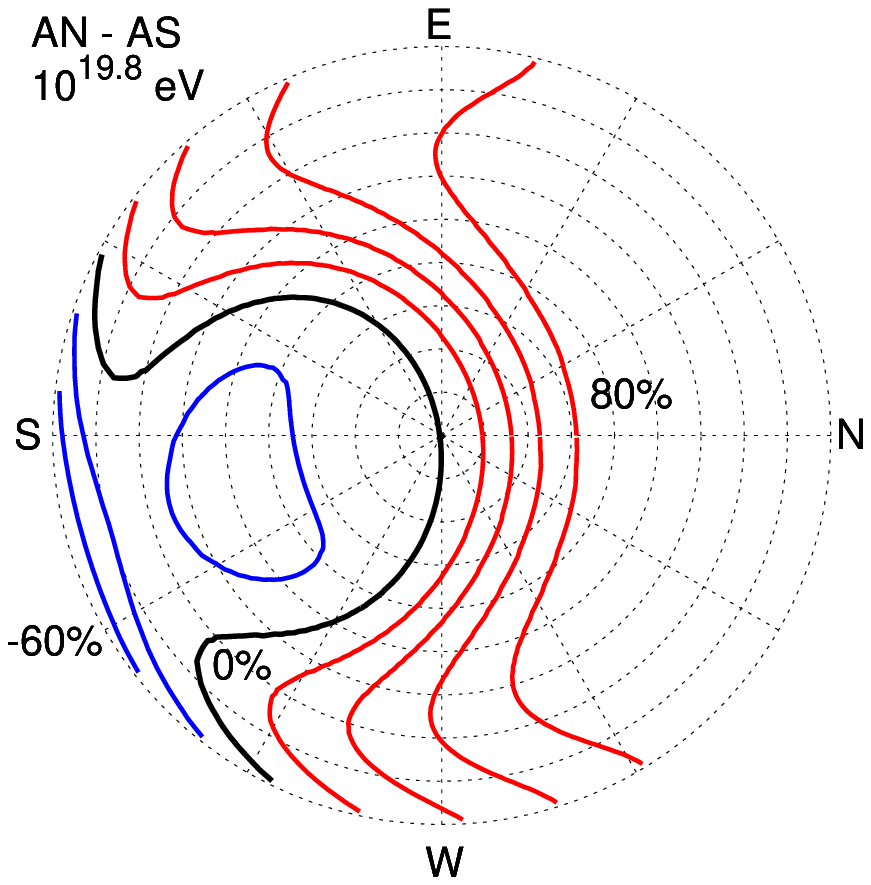}
\includegraphics[height=4.8cm,angle=0]{a_conv_diff_20.0.eps}
\caption{
Sky maps of conversion probabilities at $\lg(E/$eV)=19.8 (left)
and 20.0 (right) for Auger North (top), Auger South (middle), and
the difference (bottom); 
see also Figures~\ref{fig-pc} and ~\ref{fig-app1}
for further explanations.
}
\label{fig-app2}
\end{center}
\end{figure}
\begin{figure}[h]
\begin{center}
\includegraphics[height=4.8cm,angle=0]{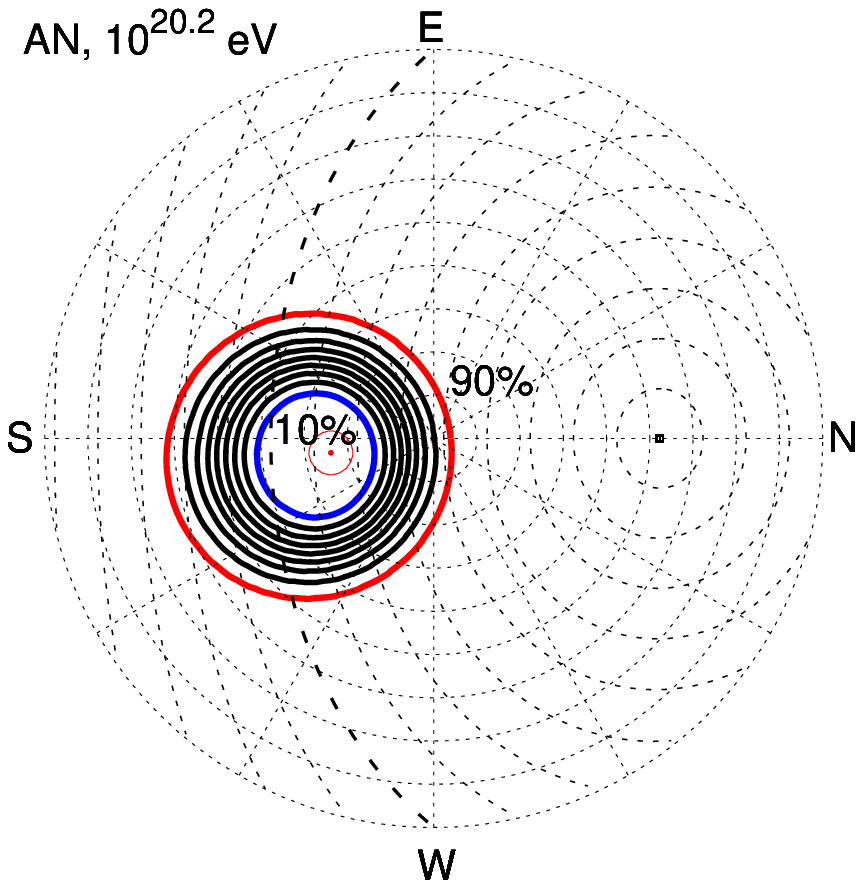}
\includegraphics[height=4.8cm,angle=0]{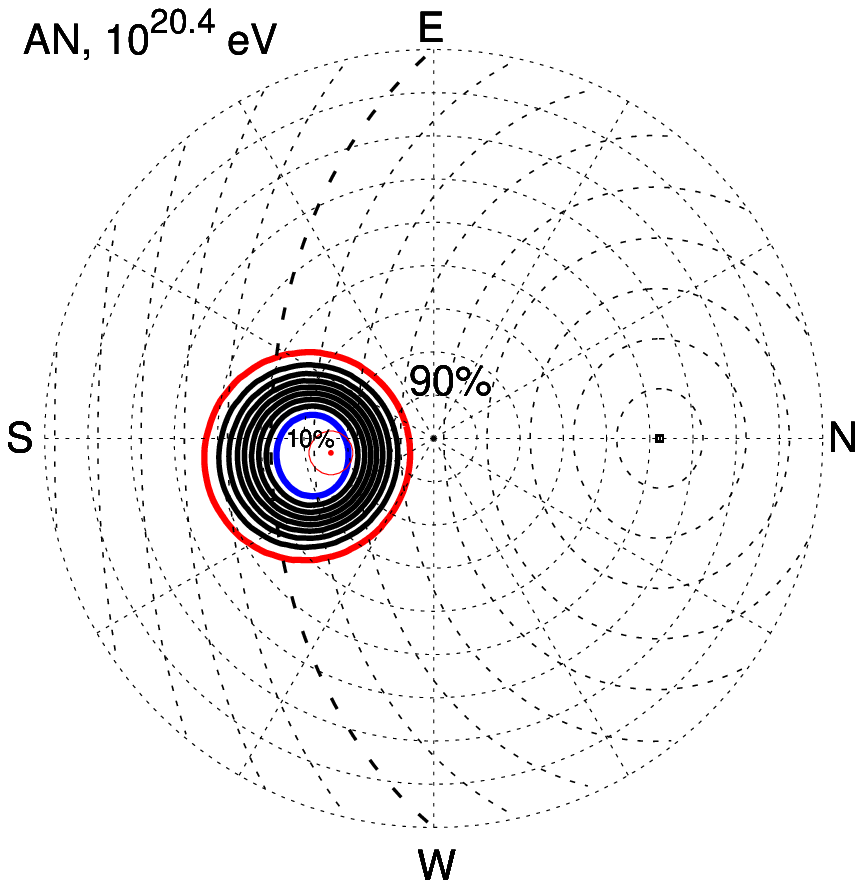}
\includegraphics[height=4.8cm,angle=0]{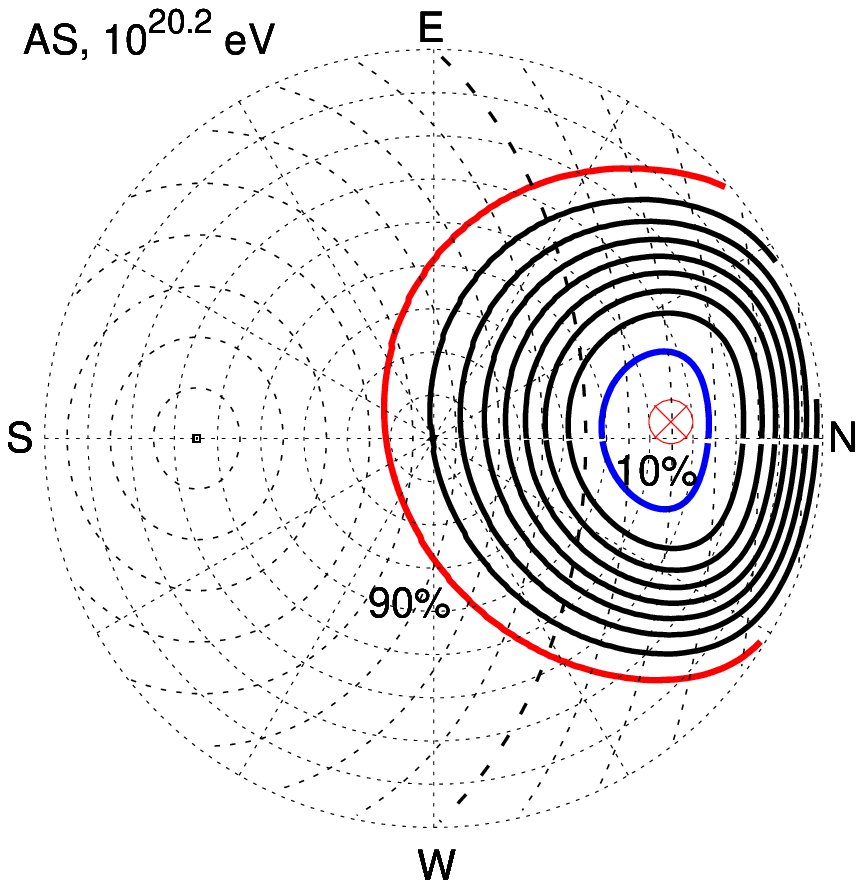}
\includegraphics[height=4.8cm,angle=0]{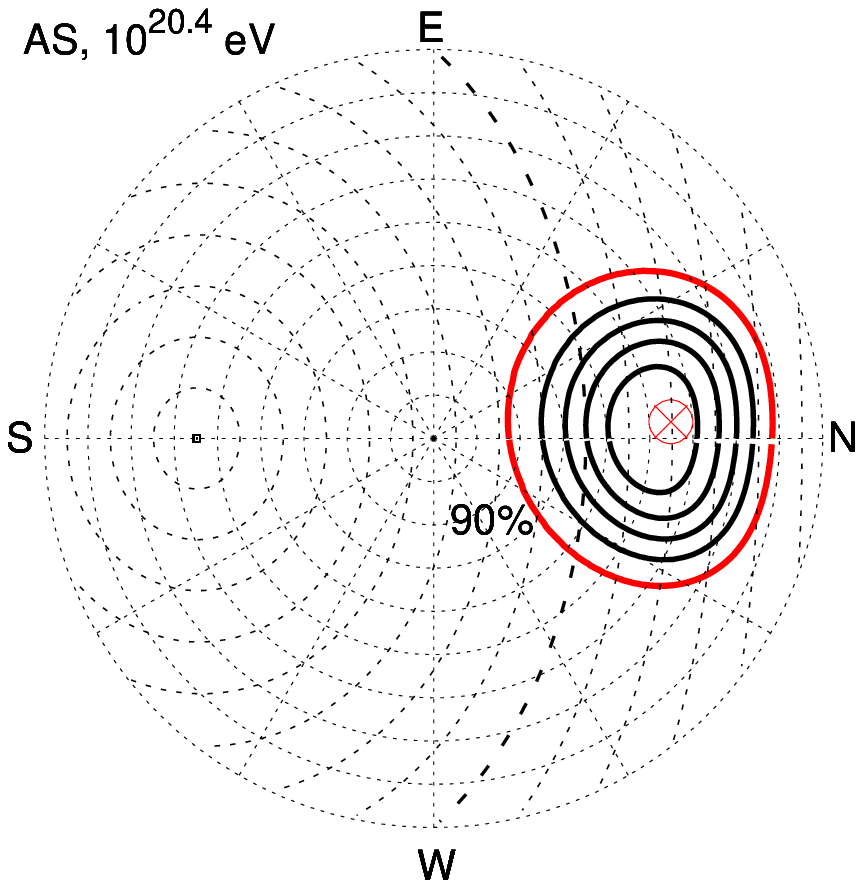}
\includegraphics[height=4.8cm,angle=0]{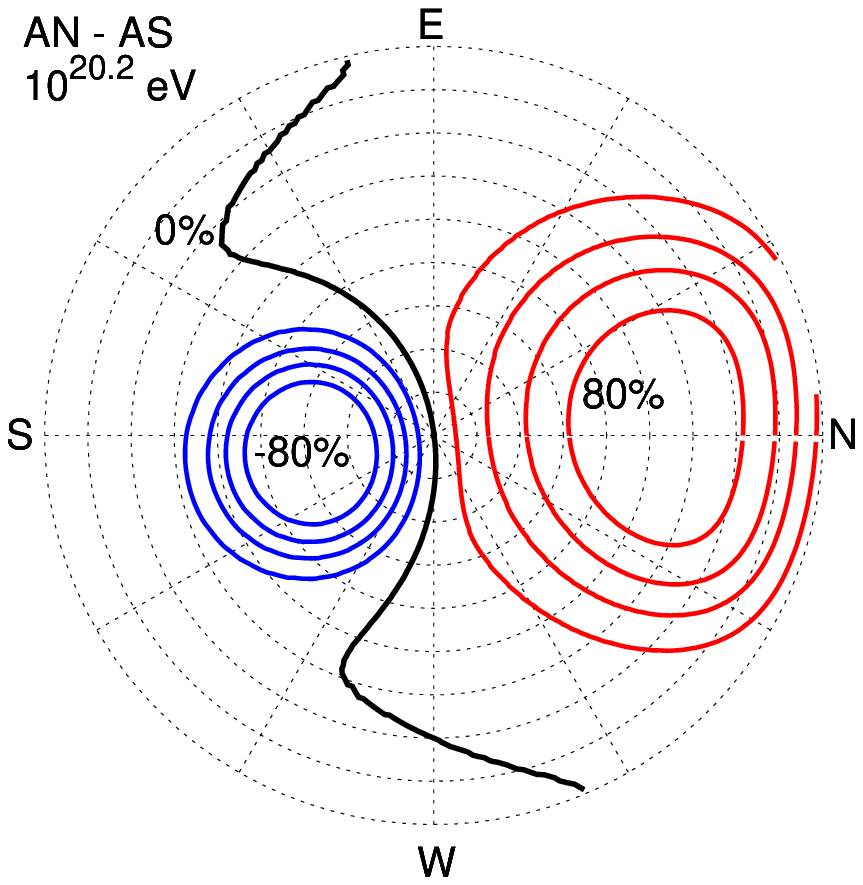}
\includegraphics[height=4.8cm,angle=0]{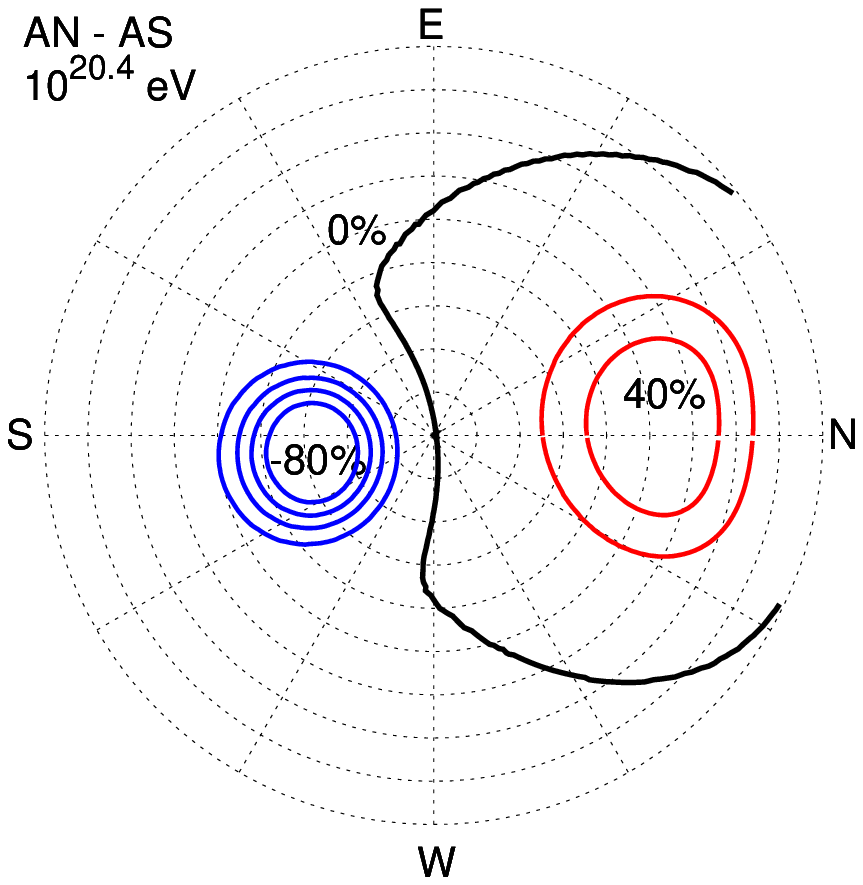}
\caption{
Sky maps of conversion probabilities at $\lg(E/$eV)=20.2 (left)
and 20.4 (right) for Auger North (top), Auger South (middle), and
the difference (bottom);
see also Figures~\ref{fig-pc} and ~\ref{fig-app1}
for further explanations.
}
\label{fig-app3}
\end{center}
\end{figure}
%
%%%%%%%%%%%%%%%%%%%%%%%%%%%%%%%%%%%%%%%%%%%%%%%%%%%%%%%%%%%%%%%%%%%%%

\end{document}